# Scan-specific Self-supervised Bayesian Deep Non-linear Inversion for Undersampled MRI Reconstruction

Andrew P. Leynes, Nikhil Deveshwar, Srikantan S. Nagarajan, *Fellow, IEEE,* and Peder E. Z. Larson, *Member, IEEE*

***Abstract*— Magnetic resonance imaging is subject to slow acquisition times due to the inherent limitations in data sampling. Recently, supervised deep learning has emerged as a promising technique for reconstructing sub-sampled MRI. However, supervised deep learning requires a large dataset of fully-sampled data. Although unsupervised or self-supervised deep learning methods have emerged to address the limitations of supervised deep learning approaches, they still require a database of images. In contrast, scan-specific deep learning methods learn and reconstruct using only the sub-sampled data from a single scan. Here, we introduce Scan-Specific Self-Supervised Bayesian Deep Non-Linear Inversion (DNLINV) that does not require an auto calibration scan region. DNLINV utilizes a Deep Image Prior-type generative modeling approach and relies on approximate Bayesian inference to regularize the deep convolutional neural network. We demonstrate our approach on several anatomies, contrasts, and sampling patterns and show improved performance over existing approaches in scan-specific calibrationless parallel imaging and compressed sensing.***

## I. INTRODUCTION

Magnetic resonance imaging is subject to slow acquisition times due to the inherent limitations in data sampling. The acquisition speed may be increased by either having gradients with higher slew rates, or by sub-sampling data below the Nyquist limit as in Parallel Imaging [1], [2] and Compressed Sensing [3]. Sub-sampling of data has been the focus of most research efforts as this allows for acceleration of scans without costly hardware changes.

Parallel imaging relies on the redundancy offered by multiple coils when sub-sampling k-space. By having multiple coils with their own local sensitivity profile, each coil image would result in different aliasing patterns. SENSE and SENSE-like approaches model the problem as a solution to a linear system of equations that unravels the aliasing artifacts to solve for the unaliased image with known coil sensitivity profiles [1], [4]. Alternatively, GRAPPA and GRAPPA-like approaches attempt to fill in the missing k-space regions using an auto calibration scan (ACS) [2]. At high acceleration rates, residual aliasing and noise amplification are encountered in these methods [5].

Compressed sensing MRI, meanwhile, relies on enforcing a sparse representation in some domain, typically using pseudo-random sub-sampling patterns that create incoherent aliasing artifacts in the sparsifying domain, and non-linear algorithms to fill in missing k-space data [6]. Compressed sensing can further accelerate MRI scans, although artifacts may be encountered that obscure anatomy and are difficult to identify [7].

Supervised deep learning has emerged as a promising technique for reconstructing sub-sampled MRI [8]. Supervised deep learning approaches learn to reconstruct images from sub-sampled data using a database of images consisting of matched pairs of sub-sampled data and fully-sampled data. These methods have shown possible acceleration factors that exceed what is possible in parallel imaging and compressed sensing [8]. However, supervised deep learning requires a large dataset of fully-sampled data.

There are scenarios where acquiring a large high-quality database of fully-sampled data may be impractical or even impossible. For example, simultaneously having ~1 mm spatial resolution at ~100 ms temporal resolution in 3D dynamic imaging is not possible with current hardware configurations and has required sophisticated reconstruction algorithms [9], [10]. Large high-quality databases are also challenging for emerging and SNR limited techniques such as non-proton imaging (e.g. $^2$H, $^{23}$Na, and $^{31}$P), chemical exchange saturation transfer (CEST) MRI, and ultra-low field MRI. Finally, obtaining large datasets including rare disease conditions are impractical due to limited availability of MRI scan data [11].

Recently, unsupervised and self-supervised deep learning have emerged to address the limitations of supervised deep learning approaches [12], [13], [14], [15], [16]. Rather than requiring fully-sampled images for supervision, these methods can utilize only sub-sampled data. Unsupervised approaches are designed to learn patterns, for example MRI priors, from the sub-sampled data, whereas self-supervised approaches explicitly create supervision signals from the sub-sampled data, which then can be used in the loss function and/or as validation data. Some recent approaches have utilized generative adversarial networks (GANs) [16], while others have directly

---

This paper was submitted for review on March 2, 2022. This project was supported in part by NIH grants R01AR078762, R01CA212148, the UCSF Graduate Research Mentorship Fellowship award, and the UCSF Cancer Center. (*Corresponding author: Peder E.Z. Larson*)

A. P. Leynes, N. Deveshwar, S. S. Nagarajan, and P. E. Z. Larson are with the Department of Radiology and Biomedical Imaging, University of California San Francisco, San Francisco, CA 94158 USA and with the UC Berkeley – UC San Francisco Joint Graduate Program in Bioengineering, Berkeley and San Francisco, CA USA (e-mail: Peder.Larson@ucsf.edu).

used model-based approaches [12], [13], [14], [15]. These methods have been demonstrated to be superior to "classical" parallel imaging and compressed sensing and competitive performance compared to fully supervised methods.

Within unsupervised and self-supervised deep learning approaches, scan-specific deep learning methods learn and reconstruct using only the sub-sampled data from a single scan (N=1). The major benefit of a scan-specific approach is that it removes the need for a training database and they operate at the lower bound of dataset size. (In practice, having additional data for testing and choose regularization parameters is often required [17], [18], similar to classical methods.) Scan-specific approaches promise to be more generalizable across hardware configurations, contrast mechanisms, pulse sequences, or data sampling schemes, much like classical reconstruction methods.

Current deep learning scan-specific methods including training a neural network to learn how to interpolate the sub-sampled regions of k-space from the fully-sampled auto-calibration scan region (like GRAPPA) [19], [20], [21], [22], [23]. A variation of Deep Decoder has also been used, either directly estimating individual coil images or sensitivity maps and a coil combined image [17].

Calibrationless methods allow for the greatest flexibility in k-space sampling by eliminating the need for an autocalibration region, reducing sampling restrictions as well as providing more options for pulse sequence designs. This relies on the principles of jointly estimating image content and coil sensitivity profiles [24], [25], [26], and often low-rank matrix completion methods have been used to solve this reconstruction problem [27], [28], [29], [30]. Supervised deep learning [31], [32] and the deep image prior method [33] has been combined with joint estimation, although these primarily test on data with an auto-calibration region. Scan-specific methods using neural networks to interpolate sub-sampled k-space have also been extended to calibrationless sampling [22].

Of relevance to this work, Bayesian methods have shown promise integrated in deep learning-based MRI reconstruction methods, including within the loss functions to improve performance [34] as well as using diffusion models to provide estimates of reconstruction uncertainty [35].

We introduce Scan-Specific Self-Supervised Bayesian Deep Non-Linear Inversion (DNLINV) for under-sampled MRI reconstruction. Our contribution is two-fold. First, we utilize a network architecture that allows for the joint estimation of image content and coil sensitivity profiles. Second, we use Bayesian deep learning to statistically regularize the network and reduce overfitting to noise. We demonstrate our approach on several anatomies, sampling patterns, and show superiority over existing approaches in scan-specific calibrationless parallel imaging and compressed sensing.

## II. THEORY

### A. Generative modeling

We define $x$ as the underlying true image. Then, using a SENSE signal model [1], the acquired coil-by-coil k-space data can be written as a linear equation:

$$y_i = SFC_i x + n \mid 1 \leq i \leq N \quad (1)$$

where N is the number of coils, $y_i$ is the k-space data for the i-th coil, $C_i$ is the coil sensitivity matrix for the i-th coil, $F$ is the Fourier transform operator, $S$ is the k-space sampling operator, and $x$ is the underlying MR image. The coil sensitivities are acquired either through a calibration scan [1] or by estimating the sensitivities from an autocalibrating scan region [4]. The optimization problem then reduces to an inverse problem of obtaining $x$ from the k-space measurements $y_i$.

Using sparsity-based regularization for reconstruction as in compressed sensing leads to the following optimization problem:

$$\hat{x} = \underset{x}{\operatorname{argmin}} \frac{1}{2} \|y - Ax\|_2^2 + \lambda_1 \|\Psi x\|_1 \quad (2)$$

$$y = [y_1 \ y_2 \ \dots \ y_N]^T \quad (3)$$

$$A = [SFC_1 \ SFC_2 \ \dots \ SFC_N]^T \quad (4)$$

Where $y$ is the complete measured k-space data from all coils, $A$ is the forward operator, $x$ is the image to be estimated, $\Psi$ is a sparsifying transformation, and $\lambda_1$ is a regularization parameter. This enforces that the image $x$ would be sparse in the transformed domain defined by $\Psi$. Common choices for $\Psi$ are the Wavelet transform or the finite difference operator [3].

We can recast the compressed sensing model in a synthesis form that is equivalent to a generative model [36]:

$$\hat{z} = \underset{z}{\operatorname{argmin}} \frac{1}{2} \|y - A\Psi^{-1}z\|_2^2 + \lambda_1 \|z\|_1 \quad (5)$$

Where instead of solving directly for $x$, we solve for sparse coefficients $z$ and generate $x$ from the known transformation $\Psi^{-1}$:

$$\hat{x} = \Psi^{-1} \hat{z} \quad (6)$$

In general, we can write this synthesis/generative modeling problem as follows:

$$\hat{z} = \underset{z}{\operatorname{argmin}} \frac{1}{2} \|y - Ag(z)\|_2^2 + R(z) \quad (7)$$

And produce the final image estimate as follows:

$$\hat{x} = g(\hat{z}) \quad (8)$$

Where $g$ is a function that generates an image $x$ from a set of coefficients $z$, and $R$ is a chosen regularization function that enforces the desired properties of $z$. This is the form of compressed sensing with generative modeling with a previously determined generative model [37], which has been shown to provide MRI reconstructions that are robust to out-of-distribution shifts by using posterior sampling [38].

We may also want to estimate the generative function $g$ in addition to estimating $z$, with a different chosen regularization function $R$ also enforces desired properties of $g$, which has previously been applied to MRI reconstruction [39]:

$$\hat{z}, \hat{g} = \underset{z,g}{\operatorname{argmin}} \frac{1}{2} \|y - Ag(z)\|_2^2 + R(z, g) \quad (9)$$

Deep Image Prior (DIP) is a generative modeling technique that generates an image from a random noise vector using a learned convolutional neural network [40]. DIP depends on implicit regularization provided by the use of a CNN as the generative model, $g(z) = g_{CNN}(z; \theta)$, where $g_{CNN}$ is a deep convolutional neural network, and can be described as a more specific case of Eq. 9:

$$\hat{\theta} = \underset{\theta}{\operatorname{argmin}} \frac{1}{2} \|y - Ag_{CNN}(z; \theta)\|_2^2 + R(\theta) \quad (10)$$

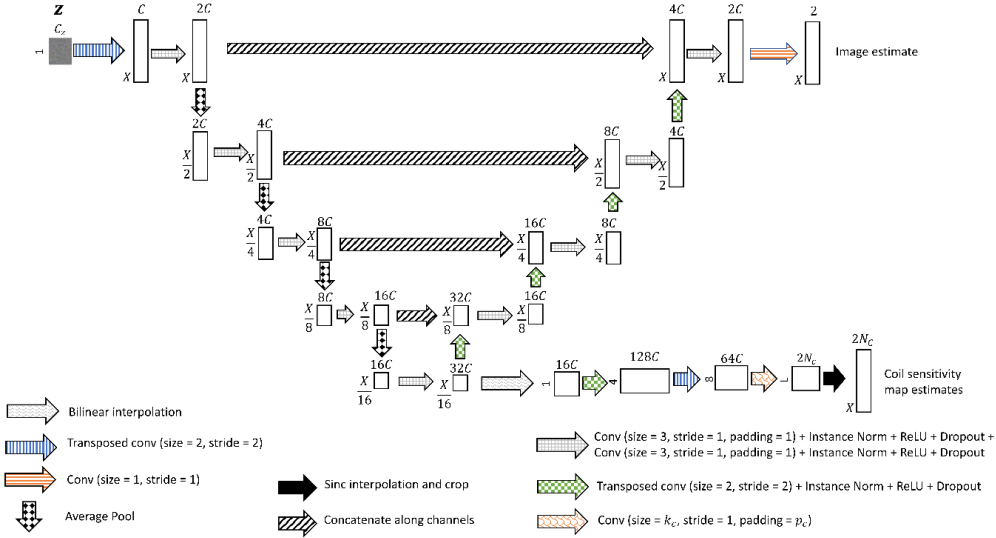

Figure 1. Deep convolutional neural network architecture used for generative modeling. The spatial size is denoted at the bottom left of each block (e.g., X) and the number of channels is denoted at the top of each block (e.g., C). This network creates both image and coil sensitivity outputs which are described in Eqns. 14 and 15. The final image estimate and each coil image have 2 channels for real and imaginary dimensions, respectively.

Where $z$ is a set of fixed random coefficients, $\theta$ are the parameters of the neural network, and $R$ is a chosen regularization function that enforces the desired properties of $\theta$. The parameters $\theta$ can be found using stochastic gradient descent using backpropagation [41] with an optimizer such as Adam [42]. The final image can be obtained similarly:

$$\hat{x} = g_{CNN}(z, \hat{\theta}) \quad (11)$$

### B. Joint estimation of the image and coil sensitivities

In calibrationless parallel imaging and compressed sensing MRI, the lack of an autocalibration scan region means that the image and coil sensitivities in eq. (1) must be solved jointly. Thus, the problem becomes non-linear. The NLINV approach [24] formulates the problem as a low-rank matrix factorization problem:

$$\hat{x}, \hat{C} = \underset{x,C}{\operatorname{argmin}} \frac{1}{2} \|y - SFCx\|_2^2 + \lambda_2(\|x\|_2^2 + \|WC\|_2^2) \quad (12)$$

$$C = [C_1 \ C_2 \ \dots \ C_N]^T \quad (13)$$

where $\lambda_2$ is a regularization parameter and $W$ is a weighting matrix penalizing high frequencies in the coil profiles. NLINV was improved by using multiple sets of coil sensitivity maps to account for model violations in ENLIVE [26], which was also shown to inherently promote low rank solutions.

For DNLINV, we reformulate the deep image prior approach by considering a new multi-task deep CNN architecture that simultaneously produces the image and coil sensitivities, shown in Figure 1. It consists of a U-net producing a coil-combined image output, denoted $g_{x_{CNN}}(z;\theta)$, with an upsampling pathway connected to the coarsest resolution feature space to produce coil sensitivities, denoted $g_{C_{CNN}}(z;\theta)$, with intended implicit low spatial frequency regularization. Note that these two outputs share the same input coefficients, $z$, and many of the model weights, $\theta$. The optimization problem for DNLINV is then:

$$\hat{\theta} = \underset{\theta}{\operatorname{argmin}} \frac{1}{2} \left\| y - SFg_{C_{CNN}}(z;\theta)g_{x_{CNN}}(z;\theta) \right\|_2^2 + R(\theta) \quad (14)$$

$$x = g_{x_{CNN}}(z;\theta), C = g_{C_{CNN}}(z;\theta) \quad (15)$$

This can also be solved using learning and inference procedures typical for deep neural networks.

### C. Bayesian inference

The DIP formulation (eq. 10) is known to overfit to noise [40], [43]. In fact, stochastic gradient descent is simply gradient descent and deep neural networks achieve zero loss under a scan-specific self-supervised regime (training $N = 1$) where cross-validation cannot be performed [44]. Other techniques such as splitting the measurements [13] could be used, but we chose to rely on Bayesian inference to combat the overfitting problem.

In contrast to regularized least-squares optimization (which is equivalent to *maximum a posteriori* (MAP) estimation with the appropriate log-prior corresponding to the regularization function $R$) performing Bayesian inference provides uncertainty estimates that has been demonstrated to reduce overfitting of the Deep Image Prior (DIP) model [43]. In particular, we utilize variational inference [45] and Monte-Carlo Dropout (MC-dropout) [46].

Reformulating the multicoil DIP problem (eq. 14) under Bayesian probabilistic modeling, we obtain the following:

$$y = SFg_{C_{CNN}}(z;\theta)g_{x_{CNN}}(z;\theta) + n \quad (16)$$

$$n \sim Normal(0, \Sigma_c) \quad (17)$$

$$p(\theta|y, z) = \frac{p(y, z|\theta)p(\theta)}{p(y, z)} \quad (18)$$

However, unlike deep image prior that only solves for the network parameters $\theta$, we also perform inference for the input coefficients to the network $z$ to solve for all unknowns in DNLINV:

$$p(\theta, z|y) = \frac{p(y|\theta, z)p(\theta, z)}{p(y)} \quad (19)$$

To use Bayesian inference we define the following priors:

$$p(z_j) = Normal(0,1) \quad (20)$$

$$p(\theta_s) = Normal(0, \lambda) \quad (21)$$

$$p(z) = \prod_{j=1}^{M} p(z_j) \quad (22)$$

$$p(\theta) = \prod_{s=1}^{S} p(\theta_s) \tag{23}$$

where $\lambda$ is the precision for the parameter $\theta_s$ from MC-dropout [46] and $S$ is the total number of network parameters.

Since directly solving for the posterior is intractable, we utilize a mean-field approximation to the posterior as follows:

$$p(\theta, z|y) \approx q(\theta|y) q(z|y) \tag{24}$$

Then, each of the approximate posteriors can be solved for independently.

To solve for the approximate posterior of the network parameters $q(\theta|y)$, we use Monte-Carlo Dropout to convert the deep convolutional neural network into a Bayesian deep convolutional network. This is an implicit method where multiple models are trained but with different dropout masks to allow for estimation of $q(\theta|y)$.

To solve for $q(z|y)$, we derive the evidence lower bound (ELBO) [45], [47] and utilize the *reparameterization trick* [45], [47]. The ELBO cost function has the following general form:

$$\mathcal{L} = E_{q(z|y)q(\theta|y)} \left[ \log \frac{p(y|z,\theta) p(z,\theta)}{q(z|y) q(\theta|y)} \right] \tag{25}$$

and we restrict $q(z)$ to be:

$$q(z_j) = Normal(\mu_{z_j}, \sigma_{z_j}), q(z) = \prod_{j=1}^{J} q(z_j) \tag{26}$$

This leads to the following cost function:

$$\begin{aligned}
\mathcal{L} &= -\frac{N}{2} (\log 2\pi + \text{logdet}(\boldsymbol{\Sigma_c})) \\
&- \frac{1}{2M} \sum_{m=1}^{M} \sum_{k=1}^{K} \Big( y_k \\
&\quad - S_k F g_C(z^{(m)}; \theta^{(m)}) g_x(z^{(m)}; \theta^{(m)}) \Big)^T \boldsymbol{\Sigma_c^{-1}} \Big( y_k \\
&\quad - S_k F g_C(z^{(m)}; \theta^{(m)}) g_x(z^{(m)}; \theta^{(m)}) \Big) - 0.5 \sum_{j=1}^{J} \sum_{m=1}^{M} (z_j^{(m)})^2 \\
&+ 0.5 \sum_{j=1}^{J} \log(\sigma_{z_j}^2) + 0.5
\end{aligned} \tag{27}$$

where $y$ is the observed k-space data, $z^{(m)}$ is $m$-th sample of the input latent code to the network shown in Fig. 1, $\theta^{(m)}$ is the $m$-th sample of the network parameters, $M$ is the number of Monte Carlo samples, $\Sigma_c$ is the coil noise covariance matrix, $g_C$ is the coil sensitivity map outputs of the network, $g_x$ is the image output of the network, $\sigma_{z_j}$ is the standard deviation of the approximate posterior $q(z_j)$, $J$ is the number of latent variables, $K$ is the number of k-space sampling points, and $N$ is the total number of k-space samples across all coils.

Finally, the optimization problem is:

$$\operatorname*{argmin}_{\mu_{z_j}, \sigma_{z_j}, \theta, \Sigma_c} \mathcal{L} \tag{28}$$

### D. Implementation details

The optimization problem in eq. 28 is solved using backpropagation stochastic gradient descent (SGD) using the Adam optimizer [42] with decoupled weight decay [48].

The nature of MC-Dropout is implicit: there is no explicit realization of $q(\theta|y)$ nor explicit cost function for it. Furthermore, we utilize decoupled weight decay [48] that places the gradient descent update corresponding to the Gaussian prior on the network parameters to the optimization update and thus does not appear in the cost function. Therefore, in the implementation of the optimization, we proceed using only $\mathcal{L}$ as the cost function but each step of SGD is performed with a different dropout mask (as in MC-Dropout). The network parameters are updated at each SGD step including the decoupled weight decay updates.

The latent code distribution parameters, $\mu_{z_j}, \sigma_{z_j}$, are updated via backpropagation using ELBO and the reparameterization described previously.

The coil noise covariance matrix $\Sigma_C$ is decomposed to a lower triangular Cholesky factorization:

$$\Sigma_c = LL^T \tag{29}$$

and $L$ is estimated instead to guarantee that $\Sigma_c$ is positive semi-definite.

The algorithm requires determining the hyperparameters of learning rate, dropout rate, weight decay, and number of iterations. Thus far, these were empirically chosen based on convergence behavior, PSNR/SSIM metrics, and image quality using a subset of test cases during algorithm development. We allowed for different hyperparameters to be used for each dataset described below (ENLIVE brain, ENLIVE knee, and fastMRI knee), but these were fixed for all samples within each datasets.

The computational complexity is like that of deep learning training. The optimization is based stochastic gradient descent (SGD) that has less than linear convergence. Using reparameterization gradient variational inference [45] reduces the variance of the gradient estimates. Having a larger number of Monte-Carlo samples per SGD step reduces the gradient estimate variance and accelerates convergence at the cost of requiring more memory per SGD step.

The final image and coil sensitivity estimates are then constructed through Monte-Carlo sampling of the expectation:

$$\hat{x} = E_{q(z)q(\theta)}[g_x(z;\theta)] \approx \frac{1}{M} \sum_{m=1}^{M} g_x(z^{(m)}; \theta^{(m)}) \tag{30}$$

$$\hat{C} = E_{q(z)q(\theta)}[g_C(z;\theta)] \approx \frac{1}{M} \sum_{m=1}^{M} g_C(z^{(m)}; \theta^{(m)}) \tag{31}$$

## III. NUMERICAL EXPERIMENTS

In the spirit of reproducible research, we provide our code and scripts to reproduce all experiments in this paper at https://www.github.com/LarsonLab/dnlinv . All reconstruction parameters of the experiments can be found in the code and the results shown in this paper are immediately provided in the repository.

### A. Evaluation metrics

The evaluation metrics used were peak signal-to-noise ratio (PSNR) and structural similarity index (SSIM). PSNR was calculated as:

$$PSNR = -20 \log I_{max} + 10 \log MSE \tag{33}$$

$$MSE = \sum_{r=1}^{R} (x_r - x_{ref,r})^2 \quad (32)$$

where $x_r$ is the $r$-th voxel of the reconstructed sum-of-squares image, $x_{ref,r}$ is the $r$-th voxel of the reference fully-sampled sum-of-squares image, and $R$ is the total number of voxels in an image, while SSIM was calculated in the same manner as in the fastMRI challenge [49].

### B. 1d denoising

As an initial test of the approach, including convergence and noise properties, we compare DIP and DNLINV, including an ablation study of the effect of the inference method (Table I) in a simple 1d denoising experiment. The data was simulated with a noise standard deviation of 0.01 and 0.1 with 8 random samples to emulate the multiple sampling provided by multiple coils. The noiseless and noisy signals are shown in Fig. 2.

Deep image prior was recommended to be used in conjunction with early stopping due to its overfitting behavior [40], [43]. Thus, we also keep track of performance vs. number of optimization steps. In addition, we include the denoising performance the maximum likelihood estimator for Gaussian noise (mean of samples).

TABLE I. List of inference methods for 1d denoising ablation experiment. This starts with deep image prior (*dip*), adds inference on the latent vector ([45], Eq. 19, *dip_mc_inference*), adds Monte Carlo dropout ([46], *dip_mc_dropout*), combines the two (*dip_mc_z_mc_dropout*), and incorporates the ELBO cost function to estimate posterior of the latent vector (Eq. 27) with and without dropout (*dnlinv_no_dropout, dnlinv_dropout*).

| Inference method | Loss function | Inference on $z$ | Inference on $\theta$ |
|---|---|---|---|
| Deep image prior (*dip*) | MSE | Fixed $z$ | MAP $\theta$ |
| DIP MC-z (*dip_mc_inference*) | MSE | Draw from $p(z)$ | MAP $\theta$ |
| DIP MC-Dropout (*dip_mc_dropout*) | MSE | Fixed $z$ | Implicit $q(\theta)$ |
| DIP MC-z MC-Dropout (*dip_mc_z_mc_dropout*) | MSE | Draw from $p(z)$ | Implicit $q(\theta)$ |
| DNLINV w/o MC-Dropout (*dnlinv_no_dropout*) | ELBO | Estimate $q(z)$ | MAP $\theta$ |
| DNLINV (*dnlinv_dropout*) | ELBO | Estimate $q(z)$ | Implicit $q(\theta)$ |

### C. SENSE numerical phantom image reconstruction

A SENSE numerical phantom experiment was performed where coil sensitivity profiles are known. An evenly spaced sampling scheme was used to create coherent aliasing artifacts and higher uncertainty in these regions. A birdcage coil configuration with 8 channels was simulated using the SigPy python package (v0.1.16) [50]. A Shepp-Logan phantom was simulated with a maximum signal intensity of 1.0. Additive white Gaussian noise ($\sigma = 0.01$) was added to the image. Deep image prior and DNLINV reconstructions were compared.

### D. In vivo retrospective undersampling

Data from ENLIVE [26] and fastMRI [49] were used for retrospective undersampling experiments. The ENLIVE method was used for comparison with 2 sets of coil sensitivity maps. The number of iterations was selected individually for all methods and datasets, chosen to approximately maximize PSNR and SSIM.

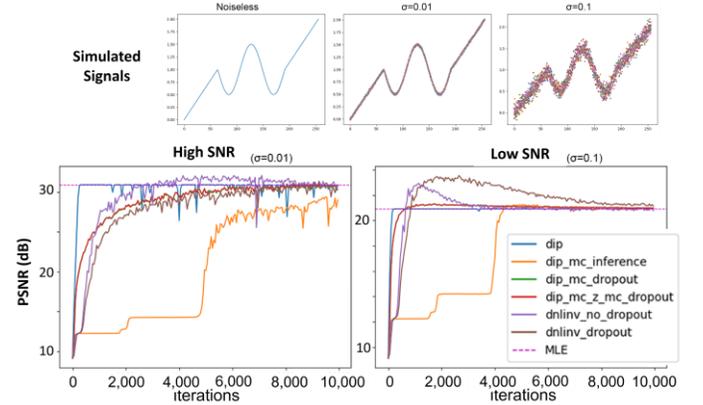

Figure 2. 1d denoising ablation experiment results demonstrating the effect of inference method (Table I) on performance and convergence in both (bottom left) high SNR and (bottom right) low SNR regimes. The noiseless and noisy signals are shown in the top row. Peak signal-to-noise ratio (PSNR) is plotted over the number of optimization iterations. The optimal maximum likelihood estimator (MLE) PSNR is shown for comparison (*dashed red line*).

#### 1) ENLIVE data

For ENLIVE data, we use both head and knee data under high acceleration factor parallel imaging and under calibrationless parallel imaging and compressed sensing sampling schemes.

Head data (3D FLASH, TR/TE = 11/4.9 ms, 3T, 32-channels) was retrospectively undersampled with a calibrationless variable-density poisson disk sampling pattern [3] at 4x, 7x, and 8.5x acceleration factors and were reconstructed using simultaneous autocalibrating and k-space estimation (SAKE) [27], ENLIVE, Deep Image Prior, and DNLINV. Additional CAIPIRINHA sampling [51] results including L2-ESPIRiT [4] are available in [52] and in our code repository.

Knee data (3D FSE, TR/TE = 1550/25 ms, 3T, 8-channels) was retrospectively undersampled with a calibrationless variable-density poisson disk sampling pattern [3] at 4x, 7x, and 8.5x acceleration factors and were reconstructed using SAKE, ENLIVE, Deep Image Prior, and DNLINV. Additional knee data used in ENLIVE were from the Stanford 3D fast-spin-echo dataset provided in mridata.org and were used to evaluate performance over a dataset.

#### 2) fastMRI data

We performed similar experiments using fastMRI knee data (2D TSE, TR/TE = 2200-3000/27-34 ms, 15-channel) [49]. Images were retrospectively undersampled with calibrationless and autocalibrated parallel imaging and compressed sensing sampling patterns (acceleration factors = [2, 3, 4]) were reconstructed with L1-ESPIRiT, SAKE, ENLIVE, DNLINV, and the pre-trained U-net fastMRI. A uniform distribution for selecting k-space lines was used for both sampling patterns.

### E. In Vivo Ablation study

Finally, we performed an ablation study using the knee ENLIVE data at 5x acceleration. Noise estimation is performed in cases where acquiring noise samples is not done and we investigate whether it is needed for accurate reconstructions. Then, we tested how the use of non-linear activations (e.g., ReLU) in Deep image prior and DNLINV impacts the results. We gradually removed portions of DNLINV and assessed the performance impact. DNLINV was compared against DNLINV without noise estimation (keeping Σ constant), deep image prior, DIP MC-z, and DNLINV with linear activations (without ReLU).

## IV. RESULTS

### A. 1d denoising experiments

The 1d denoising experiments (Fig. 2) describe how the various components of the DNLINV method contribute to the overall results. They show that, in the high SNR regime, most all methods were able to achieve a similar PSNR, while differences in performance are found in the low SNR regime. Deep image prior by itself ("dip") has the same performance as the maximum likelihood estimator for signals with Gaussian noise, but with extremely rapid convergence that is the result of the known DIP overfitting. The addition of drawing the latent variables from a prior distribution ("dip_mc_inference") deteriorated performance, while instead adding MC-Dropout to DIP ("dip_mc_dropout" and "dip_mc_z_mc_dropout") improved PSNR in the low SNR regime even when drawing latent variables from a prior distribution. Finally, creating DNLINV by incorporating the ELBO cost function to estimate posterior of the latent vector (Eq. 27) showed the best performance in the low SNR regime, and using dropout ("dnlinv_dropout") achieved the highest possible peak signal-to-noise ratio (PSNR). DNLINV also showed a broad range of iterations with high PSNR, indicating that early stopping can likely be used to achieve this performance.

### B. SENSE image reconstruction experiments

The main result for the SENSE image reconstruction experiment is shown on Fig. 3. Deep image prior was severely affected by the noise amplification (e.g. g-factor), which is expected where the parallel imaging problem is under-determined in this high undersampling (8x) regime. DNLINV, on the other hand, largely did not suffer from this noise amplification as was intended by using Bayesian inference.

Reconstructions at different number of optimization steps are shown in Fig. 4. Deep image prior immediately overfits to the noise at just 50 iterations whereas DNLINV slowly generated image content, similarly to the 1d denoising experiment. DNLINV took around 2500 iterations to reach the peak PSNR and SSIM (although residual aliasing artifacts remained) and took about 7500 iterations (in Fig. 3) to largely eliminate the aliasing artifacts. Also like in the 1d denoising experiments, DNLINV begins to overfit to noise at 10,000 iterations and eventually overfits to the noise at 20,000 iterations.

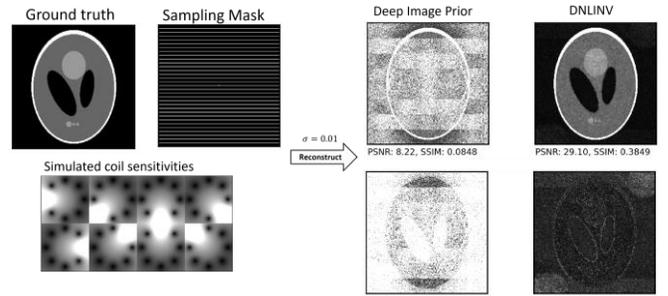

Figure 3. SENSE phantom reconstruction experiment setup with 8x undersampling. 7500 iterations were used for the image and difference images (3x scaled) result shown.

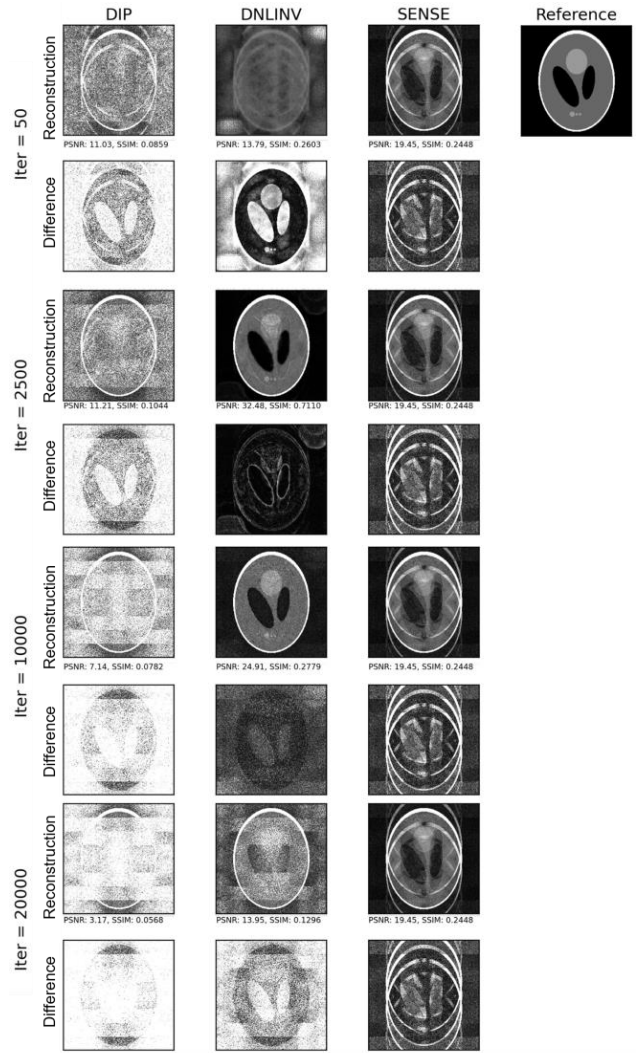

Figure 4. Phantom image reconstructions at different iterations for Deep Image Prior (1st column), Deep Non-linear Inversion (DNLINV) (2nd column), and SENSE ($\lambda_{L2} = 0.005$) (3rd column). Under each image is the absolute difference image compared to the noiseless Shepp-Logan phantom with the intensities amplified by 3x to improve visibility. Deep Image Prior shows noise overfitting due to the uncertainty in the reconstruction that is characterized by the parallel imaging g-factor. DNLINV was able to avoid this noise overfitting in earlier iterations, although also eventually overfits.

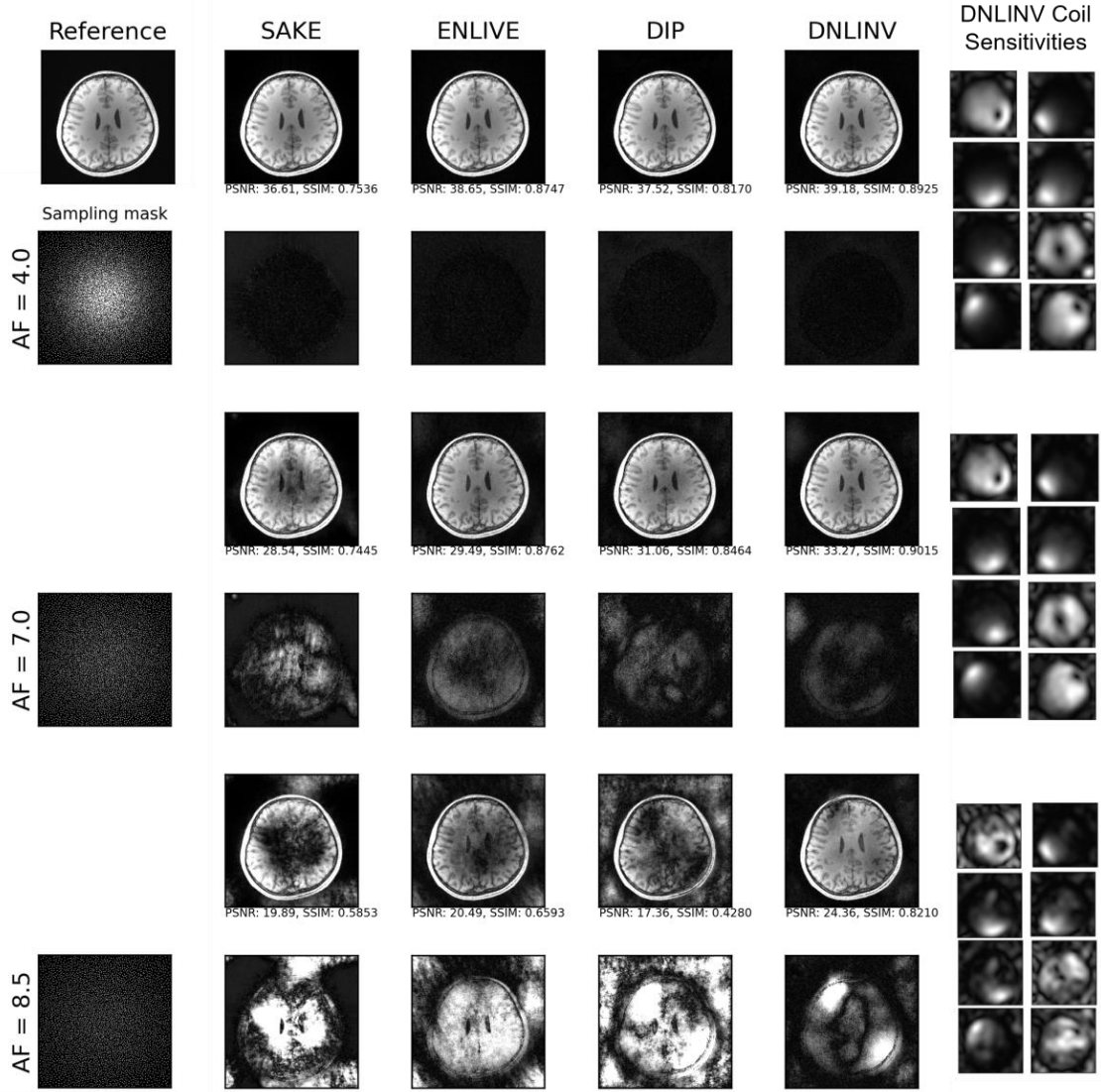

Figure 5. Head calibrationless poisson disk retrospective undersampling results comparing SAKE, ENLIVE, DIP, and DNLINV. DNLINV estimated coil sensitivities are shown for 8 out of 32 channels with the highest SNR. The absolute difference image are shown below each reconstruction and the error intensities were amplified by 3x to improve visibility. While all methods perform qualitatively similar for the lowest acceleration factor (AF) = 4.0, DNLINV showed fewer artifacts and higher PSNR/SSIM at moderate to high AF.

### C. In-vivo retrospective undersampling experiments

Figs. 5 and 6 show the main results for the calibrationless parallel imaging and compressed sensing sampling patterns. Aggregate PSNR and SSIM results over different datasets are provided in Figs. 7 and [52]. For the calibrationless 2D undersampling pattern (Fig. 5), DNLINV provided similar performance to prior reconstruction algorithms at low acceleration factors. At moderate acceleration factors, classical reconstructions (SAKE, ENLIVE) performed well but had increasing noise whereas DNLINV was able to reconstruct with similar quality to low acceleration factors. At high acceleration factors, classical reconstruction algorithms failed to properly reconstruct the images while DNLINV was still able to provide a good reconstruction. DNLINV also produced reasonable coil sensitivity profile maps without evidence of overfitting to data, although map quality degrades notably at AF=8.5. The aggregate results for 2D calibrationless sampling patterns shown in [52] demonstrate the improved performance in a dataset compared to ENLIVE and DIP.

Similar results are seen for the calibrationless 1D undersampling results in Fig. 6, although in this case DNLINV shows residual aliasing artifacts at moderate and high undersampling factors. For the aggregate dataset results in Fig. 7, DNLINV had the highest PSNR but was 2[nd] only to the pre-trained fastMRI U-net in terms of SSIM.

Results from an autocalibrating undersampling pattern in Fig. 8 show that DNLINV does not have a notable advantage in this setting. Here, the U-Net performance is greatly improved when an ACS region is present, and performs better to DNLINV in this sampling environment that is closer to what the model was trained upon.

Figure 9 shows the ablation experiment, illustrating the value of each component of the DNLINV method. While DIP shows no obvious artifacts, it suffers from clear noise amplification. As in the 1D denoising example, adding MC inference slightly degrades performance and is no better than standard DIP. DNLINV suffers from artifacts or noise amplification without ReLU non-linear activations or noise estimation, while having these methods achieves the best performance. Similar performance patterns are shown in the estimated coil sensitivity maps in [52]. This data also shows results over various iterations of all methods and illustrate rapid overfitting to noise for DIP while DNLINV has a broader region of peak PSNR results similar to the 1d denoising experiment.

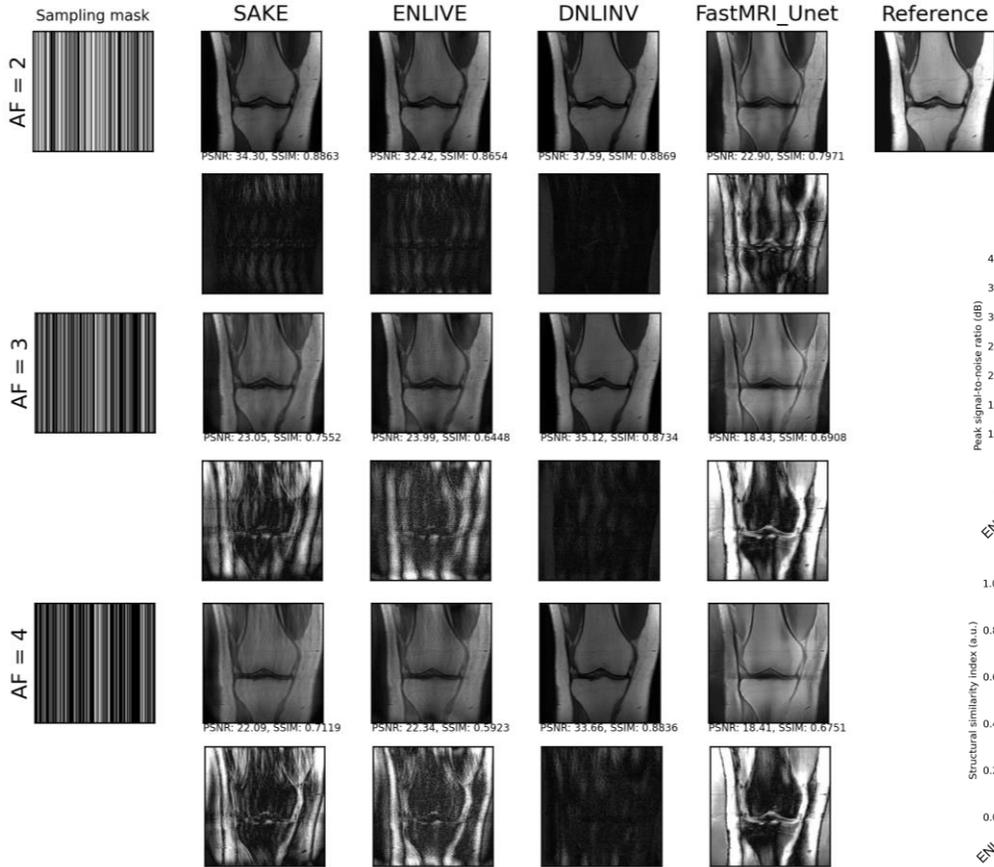

Figure 6. Knee 1d calibrationless retrospective undersampling comparing SAKE, ENLIVE, DNLINV, and the fastMRI pre-trained Unet reconstructions. The absolute difference image are shown below each reconstruction and the error intensities were amplified by 3x to improve visibility. Even at the lowest AF = 2.0, only DNLINV is the only method largely free of parallel imaging artifacts. At higher acceleration factors. While some aliasing artifacts appear at AF = 3.0 and 4.0 in DNLINV, the error is still much smaller compared to the other methods.

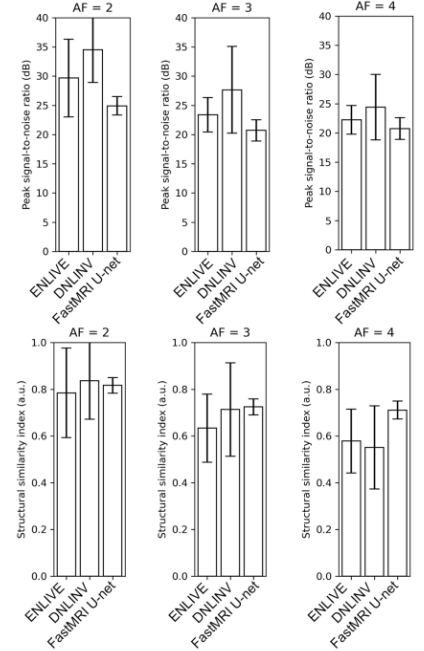

Figure 7. Aggregate reconstruction results for a calibrationless sampling pattern from the validation set of the FastMRI 3T proton-density knee dataset (n=10). Error bars correspond to 1 standard deviation of the metrics.

## V. DISCUSSION

In this paper, we show DNLINV is more robust to noise than deep image prior and similar classical methods (e.g. SAKE, NLINV, ENLIVE) and thus can reconstruct data at high acceleration factors, especially for calibrationless parallel imaging and compressed sensing. Calibrationless sampling is especially difficult due to not only undersampling of the image content, but also due to the undersampling of the low frequency k-space locations that are crucial for accurate coil sensitivity estimation. The key innovations that enabled this scan-specific, self-supervised approach are: implicit regularization with DIP-type approach, addition of upsampling path to jointly estimate the image and coil sensitivity profiles that provides further implicit regularization, and the inclusion of probabilistic Bayesian inference in the reconstruction for noise robustness that plagued prior methods.

To the best of our knowledge, there is no other scan-specific nor self-supervised deep learning method that incorporates Bayesian deep learning in reconstruction, although Bayesian deep learning has gained success in supervised learning [34], [35]. DNLINV builds off the success of DIP, and is the first of its kind to utilize Bayesian inference to minimize overfitting.

Other methods have utilized splitting the dataset into disjoint sets to separately learn the regularizer and the data fitting [13], use the autocalibration scan region for learning k-space interpolation [19], [20], [21], or rely on an architecture search to find an optimal network architecture [17].

The major component that makes DNLINV successful is the use of Bayesian evidence maximization [45], [53] using the ELBO cost function. The evidence maximization framework is what allows the network complexity to be better controlled. In DIP, only $L_p$ regularization is available to control the network complexity and since the network is overparameterized, the network still overfits [40], [43]. We demonstrated in multiple experiments the impact that Bayesian evidence maximization can have on image reconstruction.

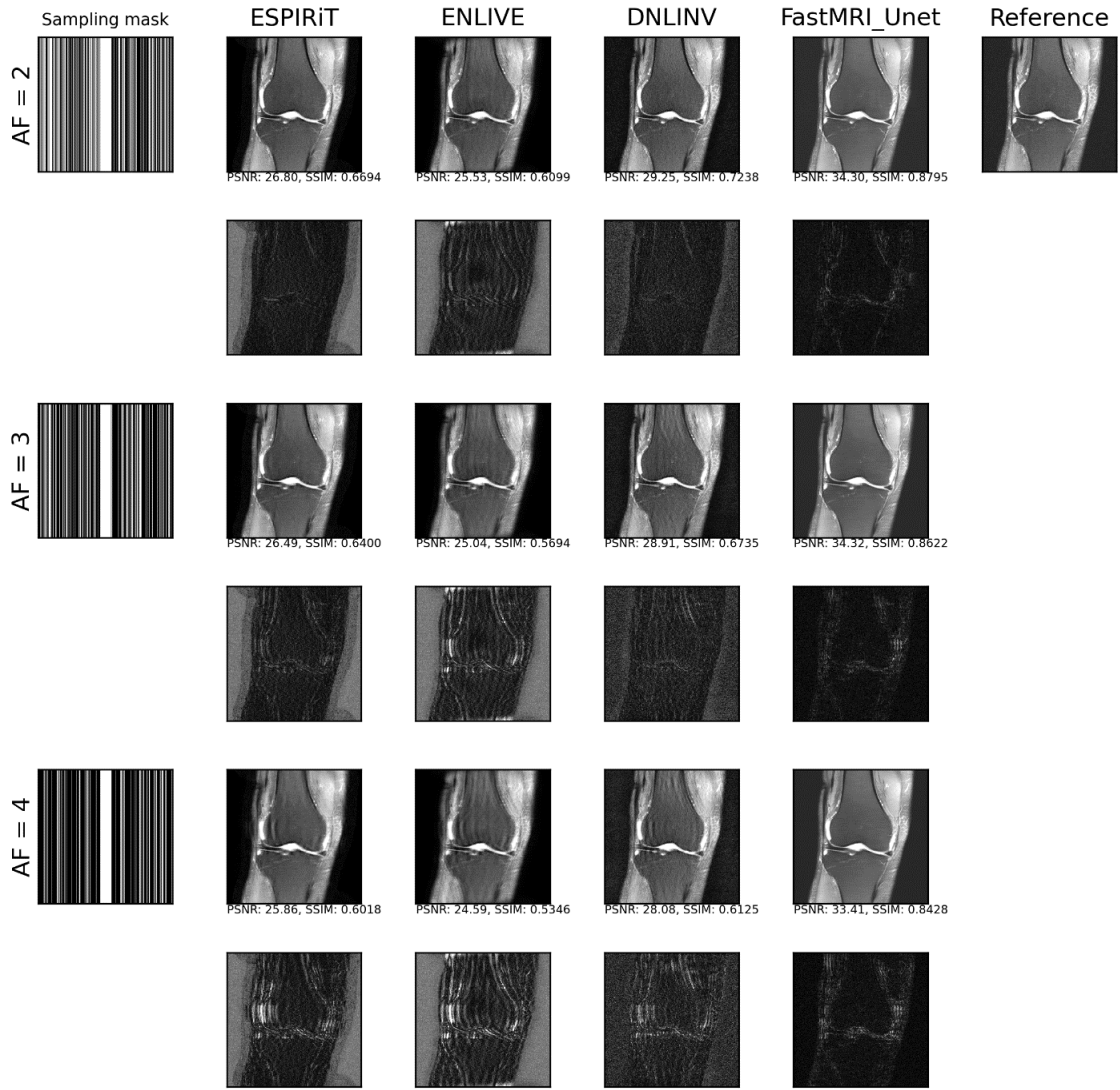

Figure 8. Knee 1d autocalibrated retrospective undersampling comparing ESPIRiT, ENLIVE, DNLINV, and the fastMRI pre-trained Unet reconstructions. The absolute difference images are shown below each reconstruction and the error intensities were amplified by 3x to improve visibility. With this sampling, ESPIRiT, ENLIVE, and DNLINV perform similarly, with noticeable aliasing artifacts at AF = 3.0 and 4.0. FastMRI_Unet outperforms the other methods in this scenario, whereas DNLINV has more advantages with calibrationless sampling as shown in other figures.

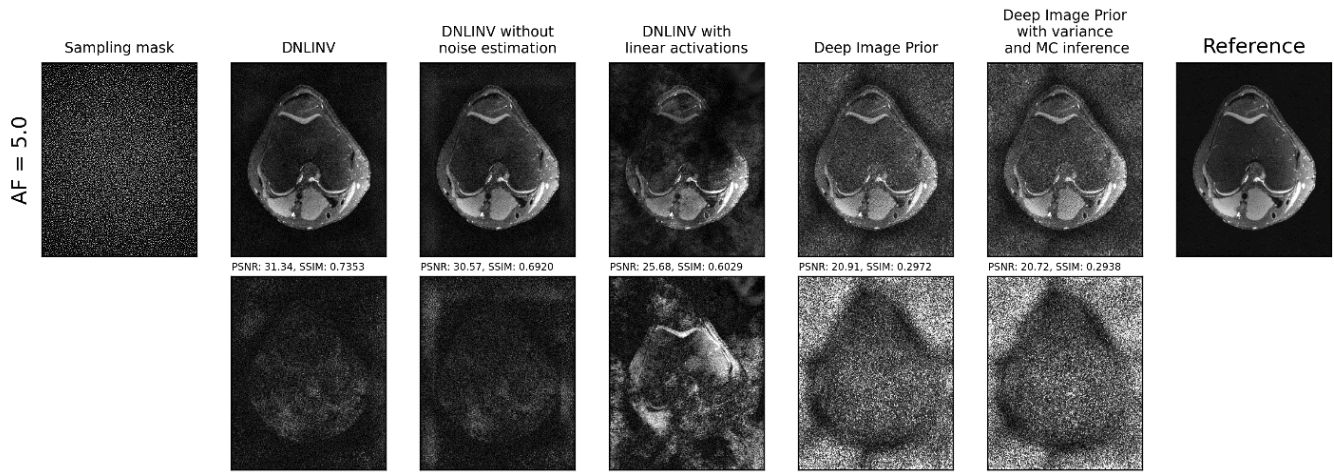

Figure 9. Ablation experiment in a knee for calibrationless retrospective 2D undersampling. The absolute difference image compared to the reference are shown below each reconstruction and the error intensities were amplified by 3x to improve visibility. Coil sensitivity maps measured during each experiment are shown in [52].

The next major component is the use of ReLU non-linear activations in the network. We demonstrated that using only linear layers lead to failure in image reconstruction (Fig. 9), showing the necessity of the non-linearities. The use of ReLU activations make the network like a multi-layer semi-non-negative convolutional sparse coding model [54].

The scan-specific and self-supervised nature of our approach allows for it to perform much like classical reconstruction algorithms and can be used for any hardware configuration, contrast, pulse sequence, or sampling pattern without any re-training. Supervised reconstruction algorithms have difficult handling test cases that are far from the training distribution. In the test example using the fastMRI data under the calibrationless context (Fig. 6), the pre-trained U-net produced significant artifacts. This is likely due to the U-net being trained on different sampling patterns that include an autocalibration region (Fig. 8). Dependence on sampling pattern and noise level is a limitation in current supervised learning approaches [55].

By eliminating the need of the autocalibration region, calibrationless parallel imaging and compressed sensing acquisitions allow for the greatest flexibility in k-space sampling patterns. The optimal sampling strategy is an active area of research [56] and maybe coupled to the reconstruction approach, and DNLINV performs well across sampling patterns. Thus far, supervised deep learning methods have been demonstrated to reconstruct calibrationless parallel imaging and compressed sensing data [31], [32], [57], [58]. The advantage of a scan-specific approach such as DNLINV, DeepDecoder [17] or LORAKI [22] is that it can be used even in the most experimental of hardware configurations and pulse sequences.

Our results show that DNLINV is able to compete with supervised approaches but only up to a moderate acceleration factors, as shown in autocalibrated sampling results (Fig. 8, [52]). Supervised approaches have the benefit of learning from a database whereas DNLINV does not, and thus would be expected to have performance improvements in well-defined, more consistent circumstances.

### A. Limitations

DNLINV still tends to overfit to noise over time, and requires early stopping, although this problem is not nearly as severe as it is for DIP. This is due to the variational approximations used, where mode-seeking and overfitting behavior is a well-known characteristic of variational Bayesian inference [45]. Using more sophisticated and computationally intensive inference methods such as through Langevin dynamics [43], expectation propagation [59], or ensemble methods [60] may be used to ultimately conquer the overfitting aspects, or by splitting up data to provide validation for early-stopping [61].

DNLINV also requires the choice of hyperparameters that are the learning rate, dropout rate, weight decay, and number of iterations. In this work, we chose the hyperparameters during algorithm development when reconstructing a small number of test cases. The learning rate, dropout rate, and weight decay were chosen by monitoring the convergence behavior. The number of iterations was selected to approximately maximize PSNR and SSIM, and is illustrated in detail in Supp. Figs. 2 and 9 [52]. Note that these results show DNLINV is relatively insensitive to the choice of number of iterations. In this initial proof of concept, we allowed the hyperparameters to vary across each dataset (ENLIVE brain, ENLIVE knee, and fastMRI knee), as they have different SNR, resolution and contrast that could influence optimal hyperparameters. However, the hyperparameters were fixed for sample within a dataset. A more principled approach to choosing this method would simplify the use of DNLINV.

There are also large computational requirements for DNLINV. The Bayesian inference used has heavy compute requirements due to the Monte Carlo approximation of the expectations and integrals and the high noise of SGD. The reconstructions performed take 15 minutes to 2 hours on GPU depending on number of iterations required for convergence. Possible ways to accelerate this process is by pre-training on a database to initialize the network to a good configuration and thus require less iterations on reconstruction (i.e., transfer learning) [17]. Alternatively, meta-learning approaches may be used to learn an optimizer that also reduces the number of gradient steps needed (i.e., learning-to-learn [62]). Finally, the use of deep convolutional neural networks has a large memory requirement. Possible ways to save on memory is with the use of more compact neural network architectures (e.g., DCGAN/DeepDecoder [17], [63]) or by reducing the overall size of the network using an architecture search [17].

Supplementary Materials for **A. P. Leynes, N. Deveshwar, S. S. Nagarajan, & P. E. Z. Larson, "Scan-specific Self-supervised Bayesian Deep Non-linear Inversion for Undersampled MRI Reconstruction"**

**Theory – Derivation of Cost Function**

Using Bayes theorem, we can write the inversion problem generally as follows:

$$p(z, \theta | y) = \frac{p(y|z, \theta) p(z, \theta)}{p(y)} \quad (1)$$

With the generative model defined as

$$y = A g_C(z; \theta) g_x(z; \theta) + n \quad (2)$$

And the additive noise $n$ is defined as

$$n \sim MultivariateNormal(0, \Sigma_c) \quad (3)$$

Where $y$ is the observed k-space data, $z$ is the input latent code to the network, $\theta$ are the network parameters, $\Sigma_c$ is the coil covariance matrix, $A$ is the forward operator, $g_C$ is the coil sensitivity map outputs of the network, and $g_x$ is the image output of the network. In the case of linearly-encoded MRI:

$$A = SF$$

where $F$ is the Fourier transform operator and $S$ is the k-space sampling operator.

The Bayesian evidence maximization framework in Chapter 3.4 of [1] poses the following optimization problem:

$$\max_M p(y|M) \quad (4)$$

Where $M$ is a model for the data. This optimization problem searches the space of all possible models and attempts to find the model that explains the data the best. In the context of DNLINV, the evidence maximization function becomes:

$$\max_{p(z,\theta|y)} p(y) = \max_{p(z,\theta|y)} E_{p(z, \theta|y)} \left[ \log \left( \frac{p(y|z, \theta) p(z, \theta)}{p(z, \theta|y)} \right) \right] \quad (5)$$

since the model $M$ is fixed to be a deep convolutional neural network. In effect, this cost function solves for the posterior distribution of the network parameters and latent code that makes the observation the most likely.

Derivation:

$$\max_{z,\theta} p(y) = \max_{z,\theta} \log p(y) \quad (6)$$

$$\log p(y) = \log p(y) \iint p(z, \theta|y) dz d\theta \quad (7) \; (Multiplication \; by \; 1 \; since \; the \; integral \; results \; to \; 1)$$

$$= \iint \log p(y) \, p(z, \theta|y) dz d\theta \quad (8)$$

($\log p(y)$ can be placed inside the integral since $y$ is not the integration variable)

$$= \iint \log\left(\frac{p(y|z,\theta)p(z,\theta)}{p(z,\theta|y)}\right) p(z,\theta|y) dz d\theta \quad (9) \text{ (by expansion of } p(y) = \frac{p(y|z,\theta)p(z,\theta)}{p(z,\theta|y)})$$

$$= E_{p(z,\theta|y)}\left[\log\left(\frac{p(y|z,\theta)p(z,\theta)}{p(z,\theta|y)}\right)\right] \quad (10) \text{ (using the definition of the expectation)}$$

The problem now shifts to solving for the posterior distribution $p(z,\theta|y)$. Given the non-linear nature of the generative model, solving for the posterior is intractable and thus, a variational approximation must be used (Chapter 10.1 in [1]). We define a factorial variational approximation (Chapter 10.1 in [1]):

$$p(z,\theta|y) \approx q(z,\theta|y) \approx q(z|y)q(\theta|y) \quad (11)$$

This considers that the approximate posterior for $z$ and $\theta$ are independent. These two approximations cause variational inference to fit to local modes of the posterior space. Now, we utilize the Kullback-Leibler (KL) divergence (Chapter 10.1 in [1]). to formulate the cost function to fit the approximate posteriors $q$ to the true posteriors $p$:

$$\min_{q(z|y),q(\theta|y)} D_{KL}(q(z|y)q(\theta|y)||p(z,\theta|y)) \quad (12)$$

To solve the above minimization problem, we derive the Evidence Lower Bound [2] under the model that uses approximate posteriors:

$$D_{KL}(q(z|y)q(\theta|y)||p(z,\theta|y)) = -E_{q(z|y)q(\theta|y)}\left[\log \frac{p(z,\theta|y)}{q(z|y)q(\theta|y)}\right] \quad (13)$$

$$= -E_{q(z|y)q(\theta|y)}\left[\log \frac{\frac{p(y|z,\theta)p(z,\theta)}{p(y)}}{q(z|y)q(\theta|y)}\right] \quad (14)$$

$$= -E_{q(z|y)q(\theta|y)}\left[\log \frac{p(y|z,\theta)p(z,\theta)}{p(y)q(z|y)q(\theta|y)}\right] \quad (15)$$

$$= E_{q(z|y)q(\theta|y)}\left[\log p(y) - \log \frac{p(y|z,\theta)p(z,\theta)}{q(z|y)q(\theta|y)}\right] \quad (16)$$

$$D_{KL}(q(z|y)q(\theta|y)||p(z,\theta|y)) = \log p(y) - E_{q(z|y)q(\theta|y)}\left[\log \frac{p(y|z,\theta)p(z,\theta)}{q(z|y)q(\theta|y)}\right] \quad (17)$$

$$D_{KL}(q(z|y)q(\theta|y)||p(z,\theta|y)) = \log p(y) + \mathcal{L} \quad (18)$$

Since $\log p(y)$ is a constant (because the data has already been observed, even though the form is unknown), then minimizing the KL-divergence is equivalent to maximizing $\mathcal{L}$, which is known as the "Evidence Lower Bound" or "Variational Free Energy". This name is derived by rearranging the equation above and noting that because the KL-divergence is a distance metric it must be $\geq 0$, then $\log p(y) \geq \mathcal{L}$. Eq. (18) may also be derived using Jensen's inequality (Section 2.2 and A.1 in [2]).

Our cost function is then:

$$\mathcal{L} = E_{q(z|y)q(\theta|y)}\left[\log\frac{p(y|z,\theta)p(z,\theta)}{q(z|y)q(\theta|y)}\right] \quad (19)$$

Expanding this:

$$E_{q(z|y)q(\theta|y)}\left[\log\frac{p(y|z,\theta)p(z,\theta)}{q(z|y)q(\theta|y)}\right]$$

$$= E_{q(z|y)q(\theta|y)}[\log p(y|z,\theta) + \log p(z,\theta) - \log q(z|y)q(\theta|y)] \quad (20)$$

$$= E_{q(z|y)q(\theta|y)}[\log p(y|z,\theta) + \log p(z,\theta)] - E_{q(z|y)q(\theta|y)}[\log q(z|y)]$$
$$- E_{q(z|y)q(\theta|y)}[\log q(\theta|y)] \quad (21)$$

$$= E_{q(z|y)q(\theta|y)}[\log p(y|z,\theta) + \log p(z,\theta)] - \iint q(z|y)q(\theta|y) \log q(z|y)\, dz d\theta$$
$$- \iint q(z|y)q(\theta|y) \log q(\theta|y)\, dz d\theta \quad (22)$$

$$= E_{q(z|y)q(\theta|y)}[\log p(y|z,\theta) + \log p(z,\theta)] - \int \left(\int q(\theta|y)d\theta\right) q(z|y) \log q(z|y)\, dz$$
$$- \int \left(\int q(z|y)dz\right) q(\theta|y) \log q(\theta|y)\, d\theta \quad (23)\ (by\ Fubini's\ theorem)$$

$$= E_{q(z|y)q(\theta|y)}[\log p(y|z,\theta) + \log p(z,\theta)] - \int q(z|y) \log q(z|y)\, dz$$
$$- \int q(\theta|y) \log q(\theta|y)\, d\theta \quad (24)$$

$$= E_{q(z|y)q(\theta|y)}[\log p(y|z,\theta) + \log p(z,\theta)] + H[q(z|y)] + H[q(\theta|y)] \quad (25)$$

Assuming that $p(z,\theta) = p(z)p(\theta)$, we then arrive at the general form of the cost function for DNLINV:

$$\mathcal{L} = E_{q(z|y)q(\theta|y)}[\log p(y|z,\theta) + \log p(z) + \log p(\theta)] + H[q(z|y)] + H[q(\theta|y)] \quad (26)$$

where $H$ is the differential entropy of a probability distribution.

**Numerically solving the inference problem**

Because the expectation $E_{q(z|y)q(\theta|y)}$ is intractable, we solve this using stochastic optimization [2]. Stochastic optimization approximates the expectation with Monte-Carlo samples by drawing from the approximate posteriors $q$. The cost function then becomes:

$$\mathcal{L} = \frac{1}{M}\sum_{m=1}^{M} \log p(y|z^{(m)},\theta^{(m)}) + \log p(z^{(m)}) + \log p(\theta^{(m)}) + H[q(z|y)] + H[q(\theta|y)] \quad (27)$$

Where $M$ is the number of Monte Carlo samples, $z^{(m)} \sim q(z)$ and $\theta^{(m)} \sim q(\theta)$ are sampled for each Monte Carlo sample.

We define the following priors:

$$p(z_j) = Normal(0,1) \tag{28}$$

$$p(\theta_l) = Normal(0, \lambda) \tag{29}$$

$$p(z) = \prod_{j=1}^{J} p(z_j) \tag{30}$$

$$p(\theta) = \prod_{s=1}^{S} p(\theta_s) \tag{31}$$

Where $J$ is the number of latent variables, $\lambda$ is the precision for the parameter $\theta_s$, and $S$ is the total number of network parameters.

Monte-Carlo dropout (MC-Dropout) is an approach that converts a deep convolutional neural network into a Bayesian deep convolutional neural network [3]. Using MC-Dropout in the deep convolutional neural network gives us $q(\theta|y)$ by training the network with dropout and sampling the different dropout masks during inference.

With $q(\theta)$ taken care of using MC-Dropout, we now solve for $q(z)$. Vanilla stochastic variational inference has high variance estimates of the ELBO, thus we utilize the "reparameterization trick" [2], [4] to reduce the variance of the gradient estimates. The reparameterization trick is performed as follows:

First, we define a factorized approximation for $q(z)$:

$$q(z_j) = Normal\left(\mu_{z_j}, \sigma_{z_j}\right) \tag{32}$$

$$q(z) = \prod_{j=1}^{J} q(z_j) \tag{33}$$

where $J$ is the number of latent variables. By defining $q(z_j)$ as a Gaussian distribution, we may reparametrize $z_j$ as follows:

$$z_j = \mu_{z_j} + \sigma_{z_j} \epsilon_j \tag{34}$$

where

$$\epsilon_j^{(m)} \sim Normal(0, 1) \tag{35}$$

Thus, $\epsilon_j$ is sampled and a sample of $z_j$ is constructed using eq. 34:

$$z_j^{(m)} = \mu_{z_j} + \sigma_{z_j} \epsilon_j^{(m)} \tag{36}$$

The reparametrization trick removes dependencies on an unknown probability distribution and exposes the parameters of the approximate distribution, $\mu_{z_j}$ and $\sigma_{z_j}$, to the backpropagation computational graph, which can then be solved for as a typical parameter in deep learning.

Plugging in the specifics that are known about the problem (eqs. 2, 3, 32) to the general form of the cost function (eq. 27), the cost function becomes:

$$\mathcal{L} = -\frac{N}{2}(\log 2\pi + \text{logdet}(\mathbf{\Sigma_c}))$$
$$-\frac{1}{2M}\sum_{m=1}^{M}\sum_{k=1}^{K}\left(y_k - S_k F g_C(z^{(m)};\theta^{(m)})g_x(z^{(m)};\theta^{(m)})\right)^T \mathbf{\Sigma_c}^{-1}\left(y_k\right.$$
$$\left. - S_k F g_C(z^{(m)};\theta^{(m)})g_x(z^{(m)};\theta^{(m)})\right) - 0.5\sum_{j=1}^{J}\sum_{m=1}^{M}(z_j^{(m)})^2 + 0.5\sum_{j=1}^{J}\log\left(\sigma_{z_j}^2\right)$$
$$+ 0.5 \quad (37)$$

where $y$ is the observed k-space data, $z^{(m)}$ is $m$-th sample of the input latent code to the network, $\theta^{(m)}$ is the $m$-th sample of the network parameters, $M$ is the number of Monte Carlo samples, $\Sigma_c$ is the coil covariance matrix, $g_C$ is the coil sensitivity map outputs of the network, $g_x$ is the image output of the network, $\sigma_{z_j}$ is the standard deviation of the approximate posterior $q(z_j)$, $J$ is the number of latent variables, $K$ is the number of k-space sampling points, and $N$ is the total number of k-space samples across all coils.

Since both approaches for solving are based on sampling and stochastic gradient descent, we may simply proceed with the optimization procedure the same as standard deep learning, with the addition of sampling $z^{(m)} \sim q(z|y)$ and dropout masks at each iteration.

To improve estimates of the ELBO, multiple Monte Carlo samples may be drawn at each optimization step. Additionally, we utilize decoupled weight decay [5] to induce $L_2$-regularization on the network parameters and does not appear in the cost function.

**Results**

Additional results are included for here for the in vivo undersampling and ablation experiments:

*In Vivo Undersampling*

ENLIVE Head data (3D FLASH, TR/TE = 11/4.9 ms, 3T, 32-channels) was retrospectively undersampled with a CAIPIRINHA sampling pattern [33] at 4x, 9x, 16, and 25x acceleration factors and were reconstructed using L2-ESPIRiT [4], ENLIVE [21], Deep Image Prior [22], and DNLINV.

fastMRI knee data (2D TSE, TR/TE = 2200-3000/27-34 ms, 15-channel) [31] were retrospectively undersampled with autocalibrated parallel imaging and compressed sensing sampling patterns (acceleration factors = [2, 3, 4]) were reconstructed using L1-ESPIRiT, ENLIVE, DNLINV, and a pre-trained U-net from the fastMRI project [31].

Sensitivity maps generated from the various methods in the knee data ablation experiment are shown.

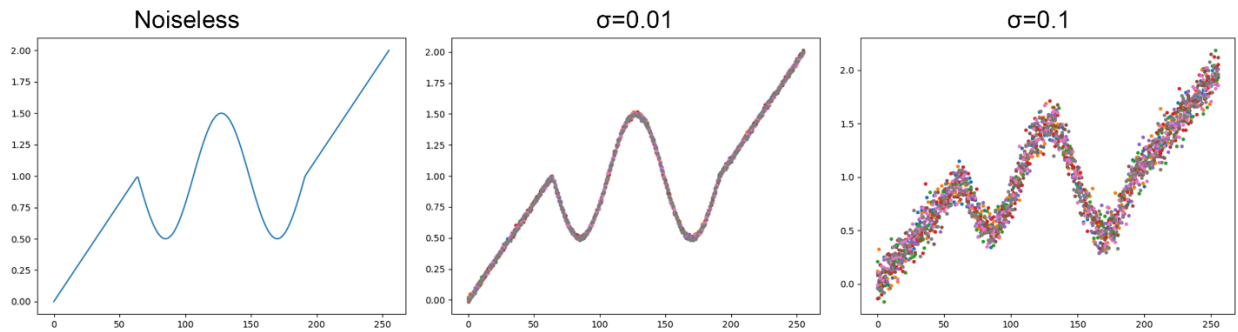

Supp. Fig. 1. (left) 1d noiseless signal, (middle) 1d signal with noise $\sigma = 0.01$, (right) 1d signal with noise $\sigma = 0.1$. There are 8 samples per location for the noisy signals.

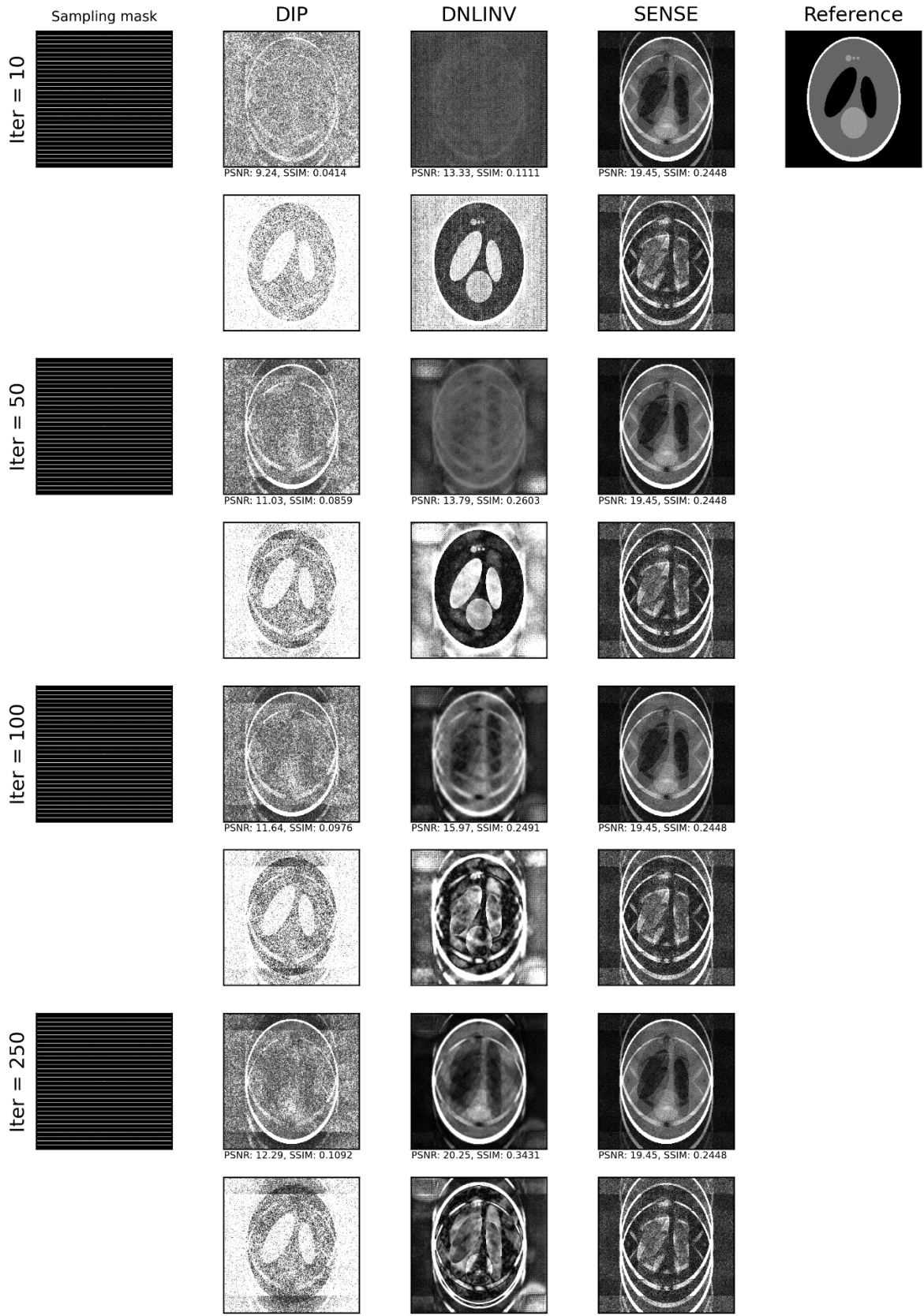

| Iter = 500 | 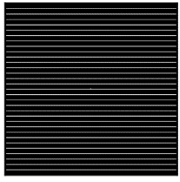 | 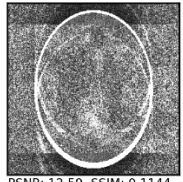 | 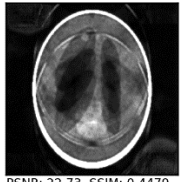 | 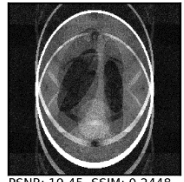 |
| --- | --- | --- | --- | --- |
| | | PSNR: 12.59, SSIM: 0.1144 | PSNR: 22.73, SSIM: 0.4479 | PSNR: 19.45, SSIM: 0.2448 |
| | | 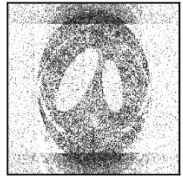 | 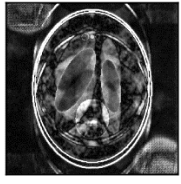 | 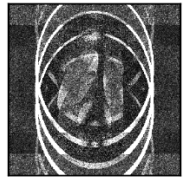 |

| Iter = 1000 | 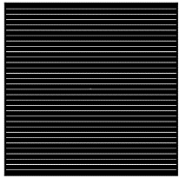 | 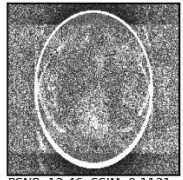 | 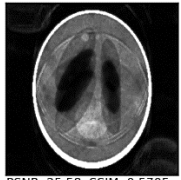 | 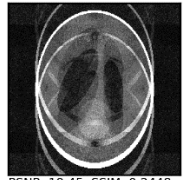 |
| --- | --- | --- | --- | --- |
| | | PSNR: 12.46, SSIM: 0.1131 | PSNR: 25.58, SSIM: 0.5705 | PSNR: 19.45, SSIM: 0.2448 |
| | | 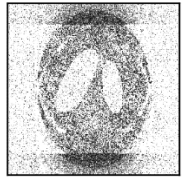 | 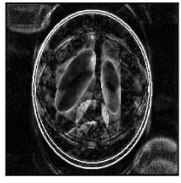 | 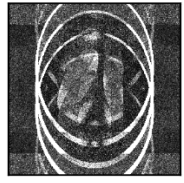 |

| Iter = 2500 | 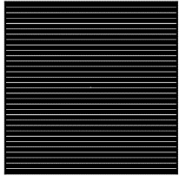 | 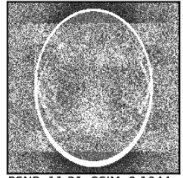 | 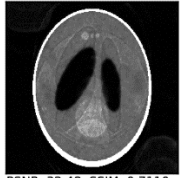 | 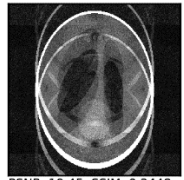 |
| --- | --- | --- | --- | --- |
| | | PSNR: 11.21, SSIM: 0.1044 | PSNR: 32.48, SSIM: 0.7110 | PSNR: 19.45, SSIM: 0.2448 |
| | | 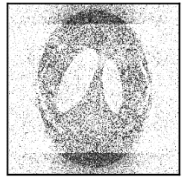 | 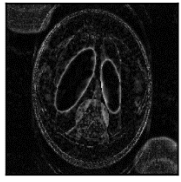 | 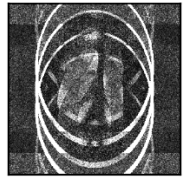 |

| Iter = 5000 | 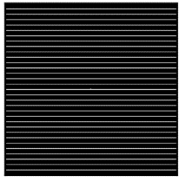 | 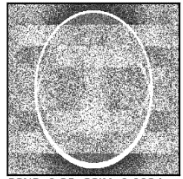 | 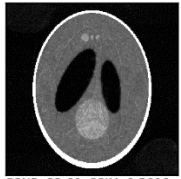 | 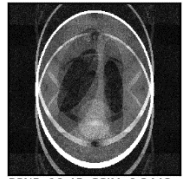 |
| --- | --- | --- | --- | --- |
| | | PSNR: 9.35, SSIM: 0.0924 | PSNR: 32.60, SSIM: 0.5606 | PSNR: 19.45, SSIM: 0.2448 |
| | | 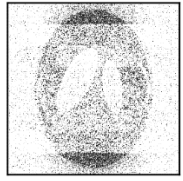 | 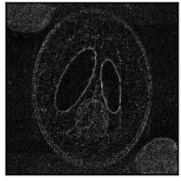 | 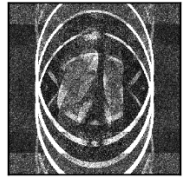 |

| Iter = 7500 | 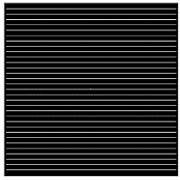 | 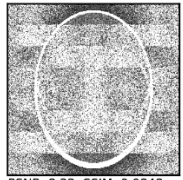 | 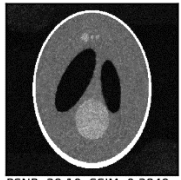 | 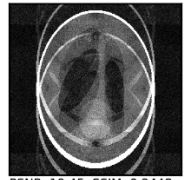 |
| --- | --- | --- | --- | --- |
| | | PSNR: 8.22, SSIM: 0.0848 | PSNR: 29.10, SSIM: 0.3849 | PSNR: 19.45, SSIM: 0.2448 |
| | | 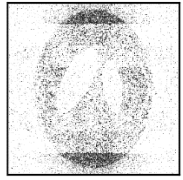 | 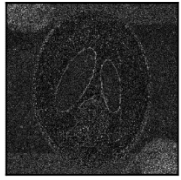 | 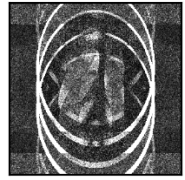 |
| Iter = 10000 | 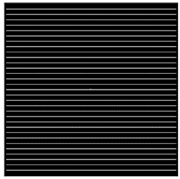 | 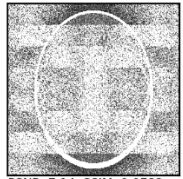 | 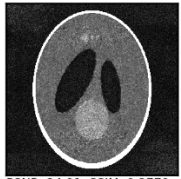 | 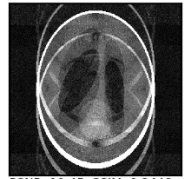 |
| | | PSNR: 7.14, SSIM: 0.0782 | PSNR: 24.91, SSIM: 0.2779 | PSNR: 19.45, SSIM: 0.2448 |
| Iter = 12500 | 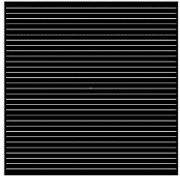 | 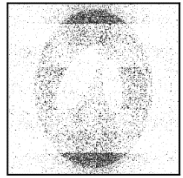 | 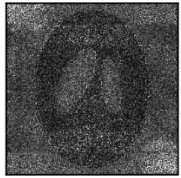 | 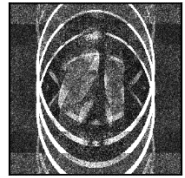 |
| | | PSNR: 6.14, SSIM: 0.0715 | PSNR: 21.80, SSIM: 0.2203 | PSNR: 19.45, SSIM: 0.2448 |
| Iter = 15000 | 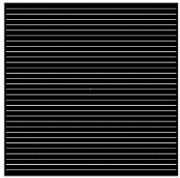 | | | |
| | | PSNR: 5.14, SSIM: 0.0653 | PSNR: 19.36, SSIM: 0.1787 | PSNR: 19.45, SSIM: 0.2448 |

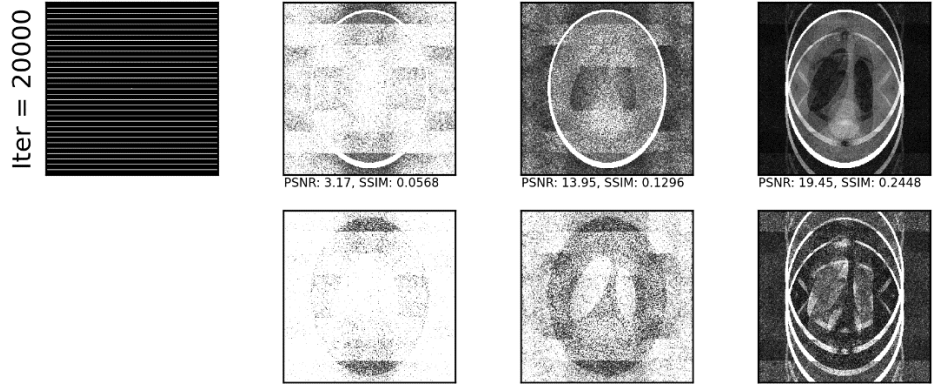

Supp. Fig. 2. Phantom image reconstructions at different iterations for Deep Image Prior (2nd column), Deep Non-linear Inversion (DNLINV) (3rd column), and SENSE ($\lambda_{L2} = 0.005$) (4th column). Under each image is the absolute difference image compared to the noiseless Shepp-Logan phantom with the intensities amplified by 3x.

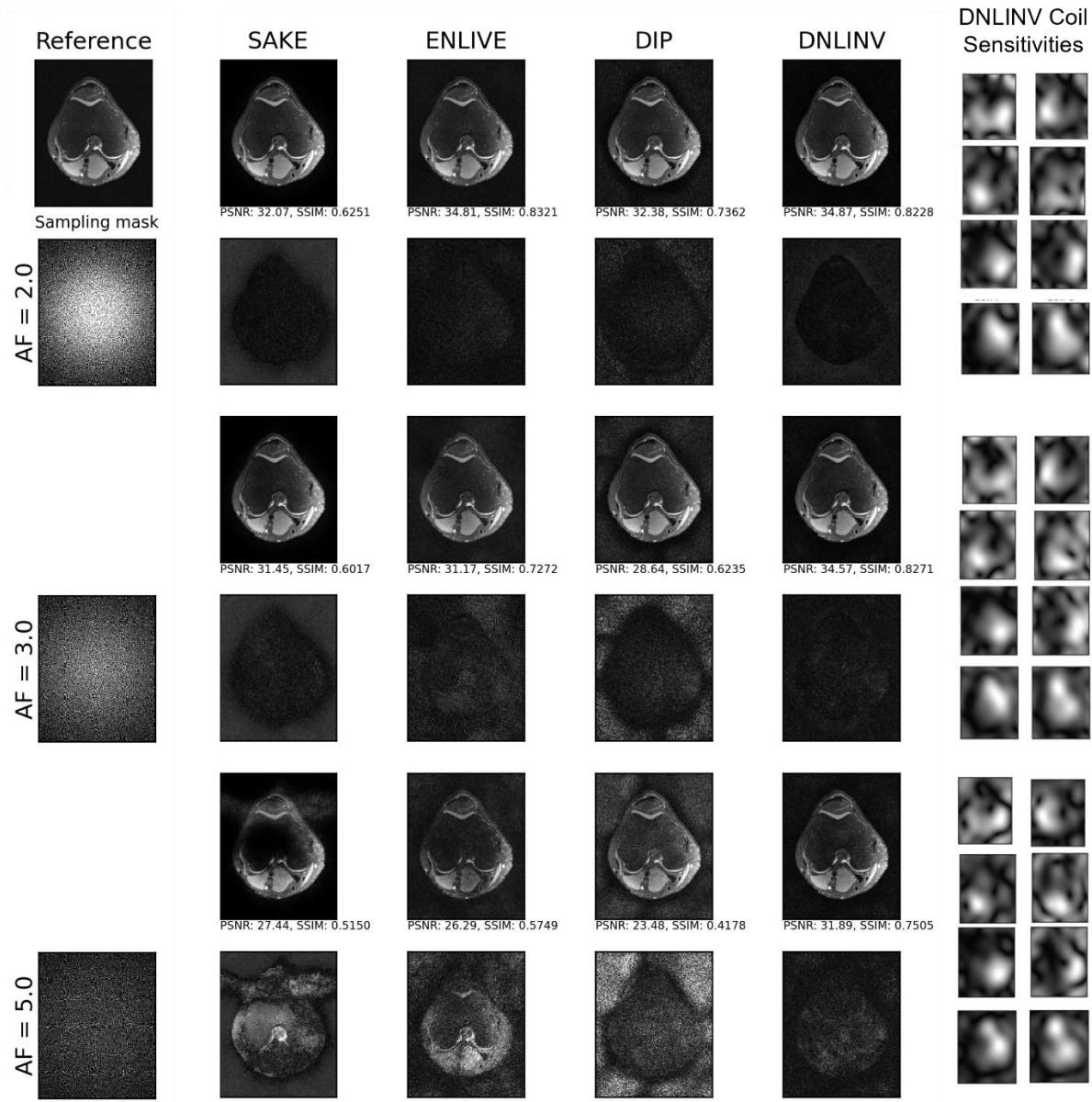

Supp. Fig. 3. Knee calibrationless poisson disk retrospective undersampling results, including sensitivity map estimations from DNLINV. The absolute difference images w.r.t. the reference image is shown under each reconstruction and the magnitudes are multiplied by 3 to better visualize the error regions.

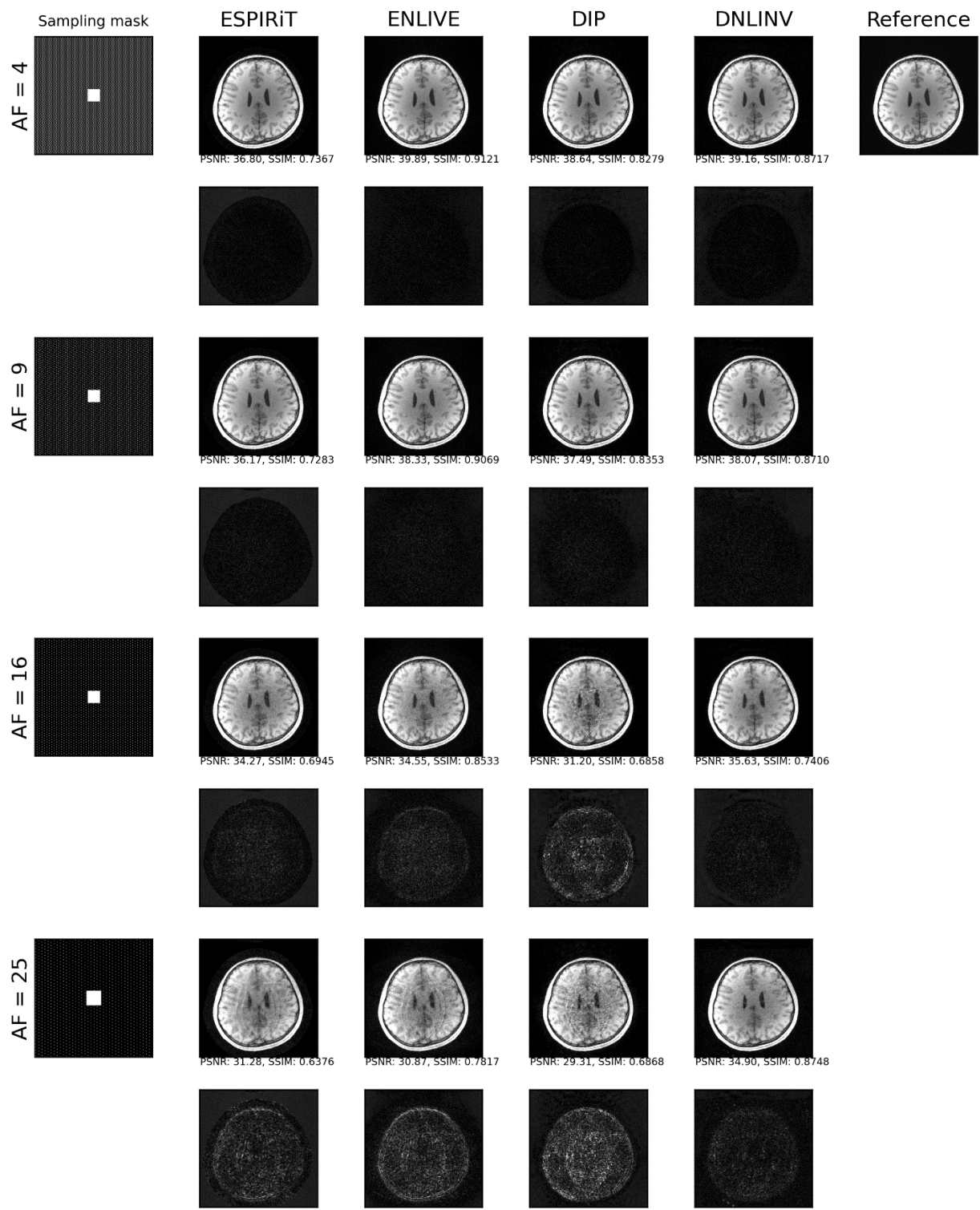

Supp. Fig. 4. Head CAIPIRINHA retrospective undersampling results. The absolute difference images w.r.t. the reference image is shown under each reconstruction and the magnitudes are multiplied by 3 to better visualize the error regions.

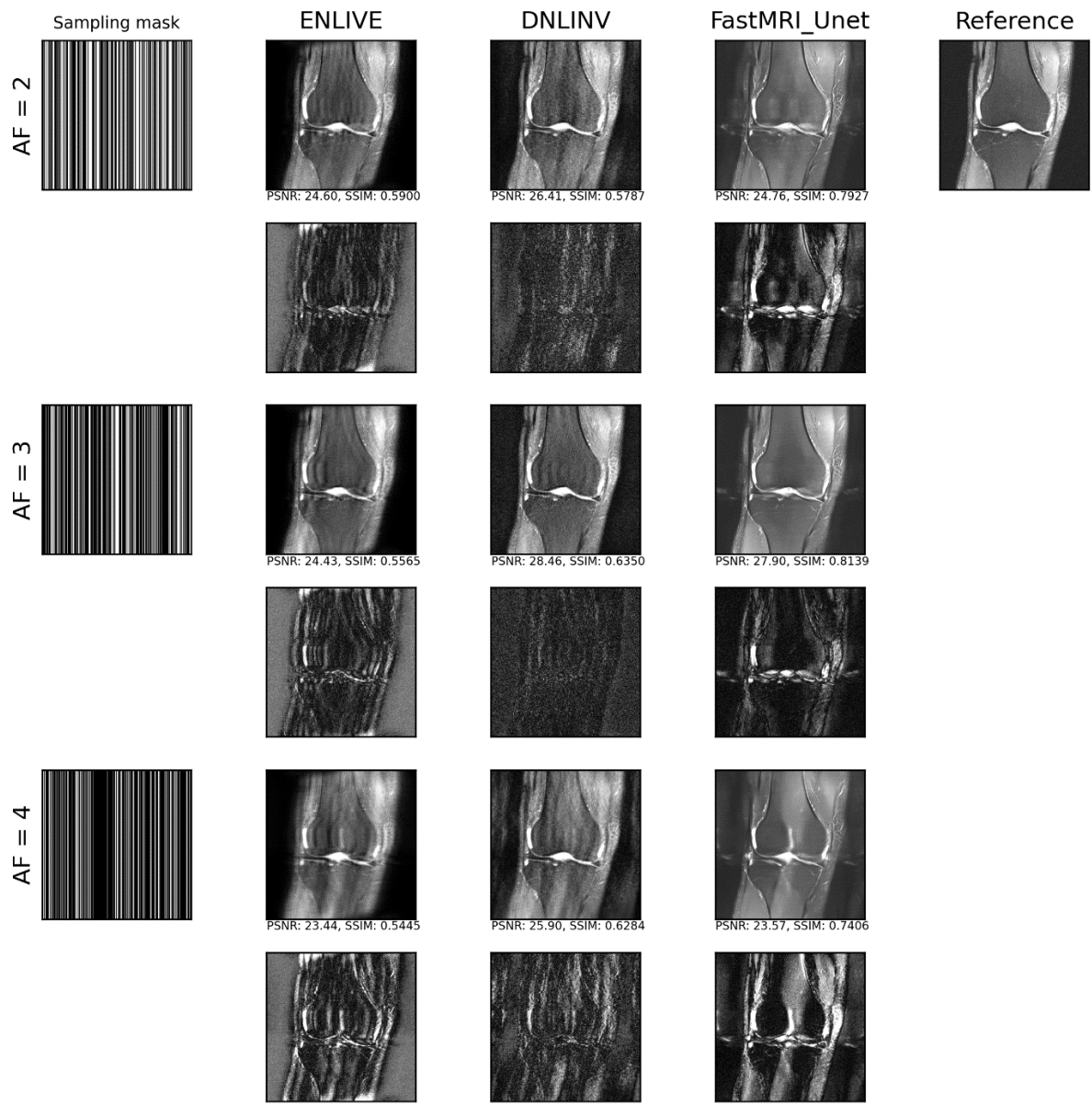

Supp. Fig. 5. Proton-density fat-suppressed knee 1d calibrationless retrospective undersampling. The absolute difference images w.r.t. the reference image is shown under each reconstruction and the magnitudes are multiplied by 3 to better visualize the error regions.

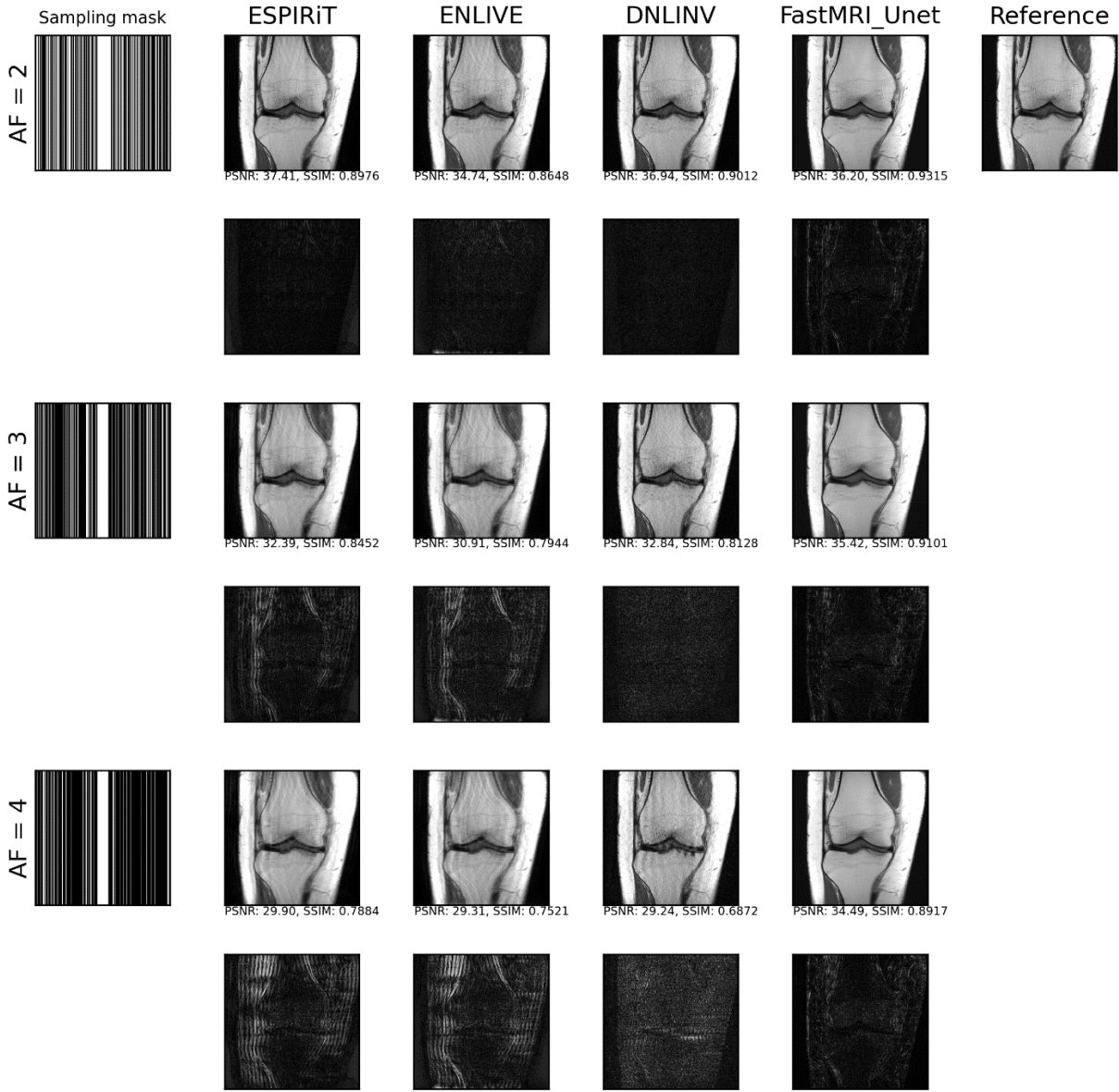

Supp. Fig. 6. Proton-density knee 1d autocalibrated retrospective undersampling. The absolute difference images w.r.t. the reference image is shown under each reconstruction and the magnitudes are multiplied by 3 to better visualize the error regions.

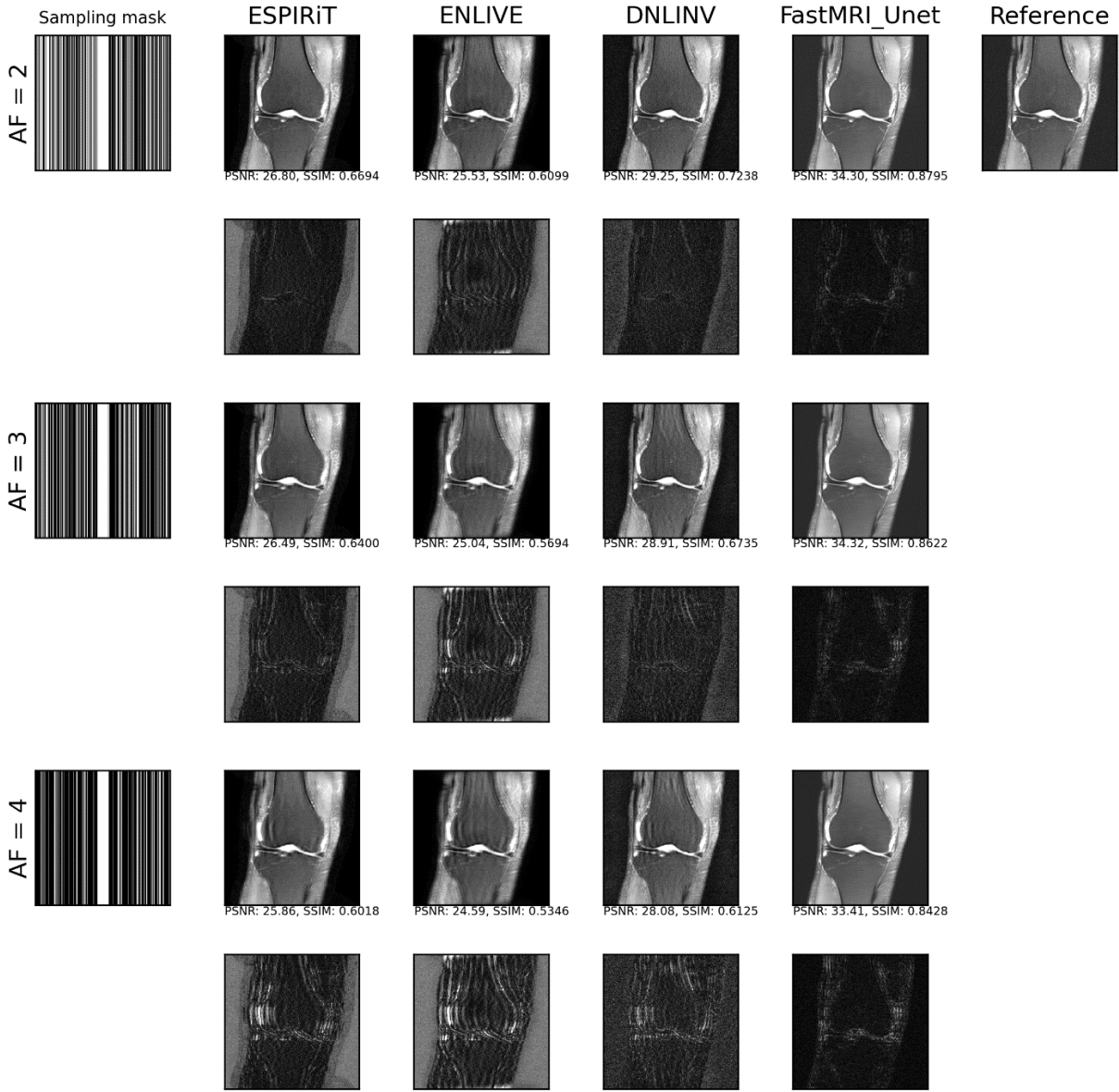

Supp. Fig. 7. Proton-density fat-suppressed knee 1d autocalibrated retrospective undersampling. The absolute difference images w.r.t. the reference image is shown under each reconstruction and the magnitudes are multiplied by 3 to better visualize the error regions.

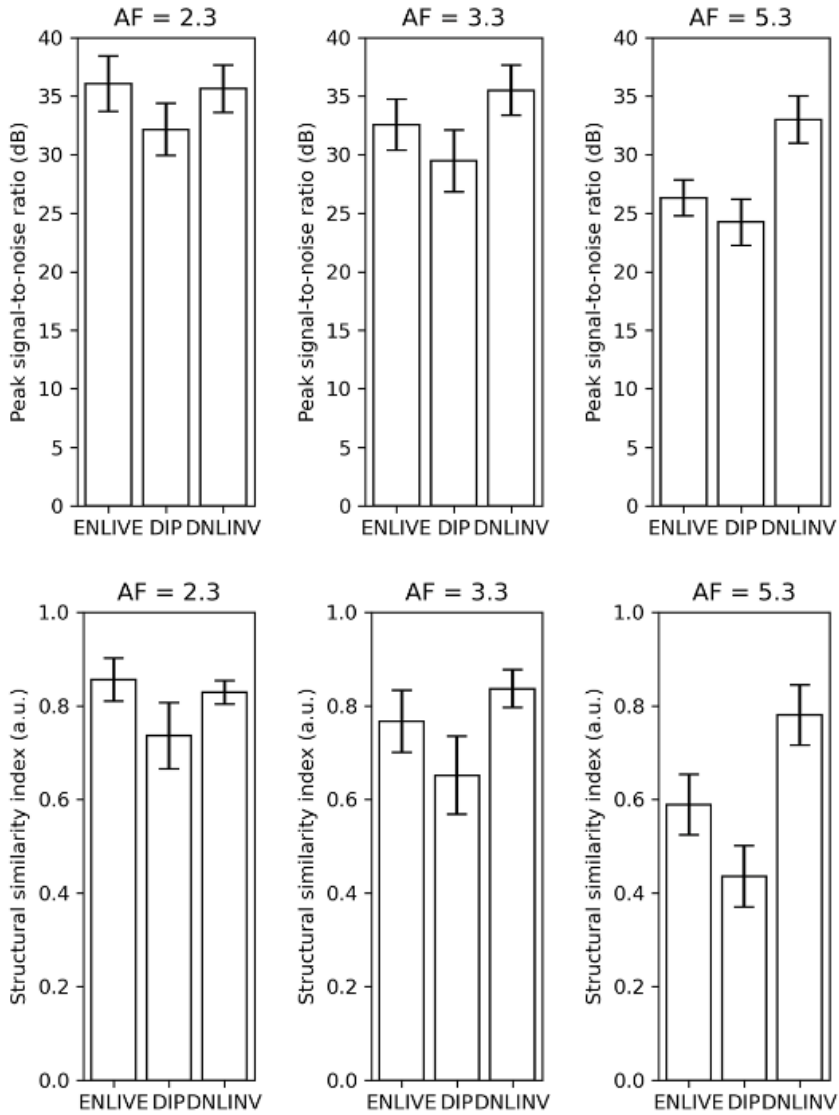

Supp. Figure 8. Aggregate reconstruction results for a calibrationless sampling pattern using the Stanford 3d knee FSE dataset from mridata.org (n=20). Error bars correspond to 1 standard deviation of the metrics.

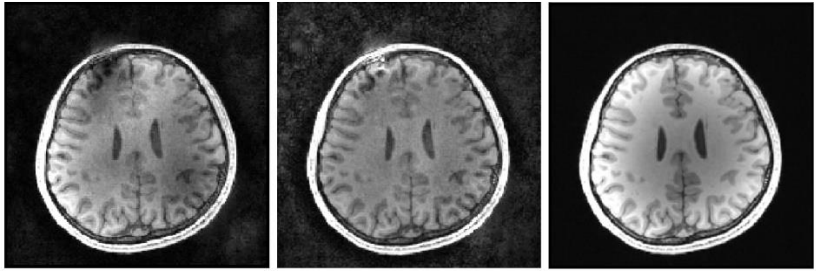

Supp. Fig. 9. DNLINV sum-of-squares image reconstruction (left), DNLINV SENSE image reconstruction (middle), and fully-sampled sum-of-squares image (right)

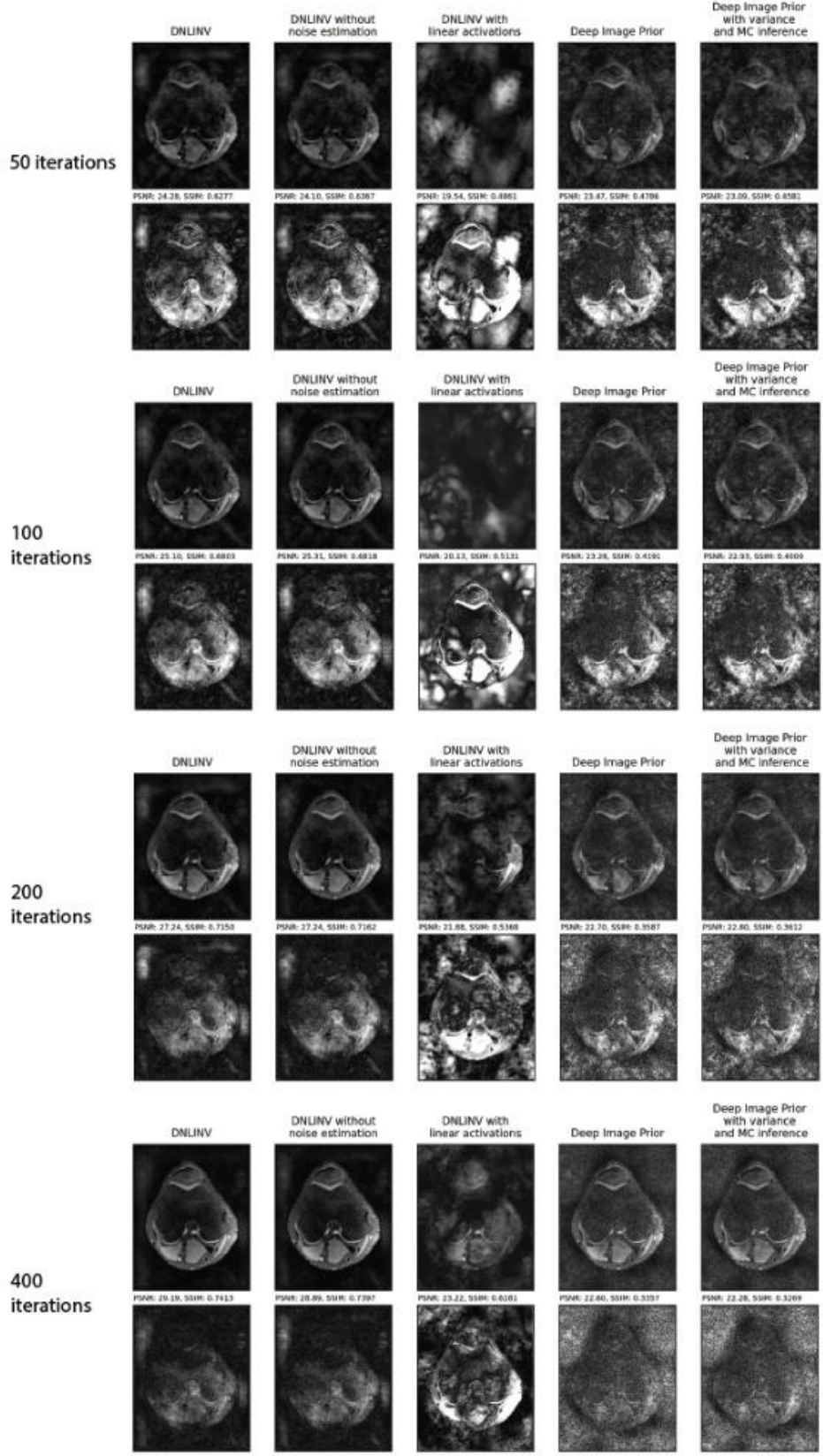

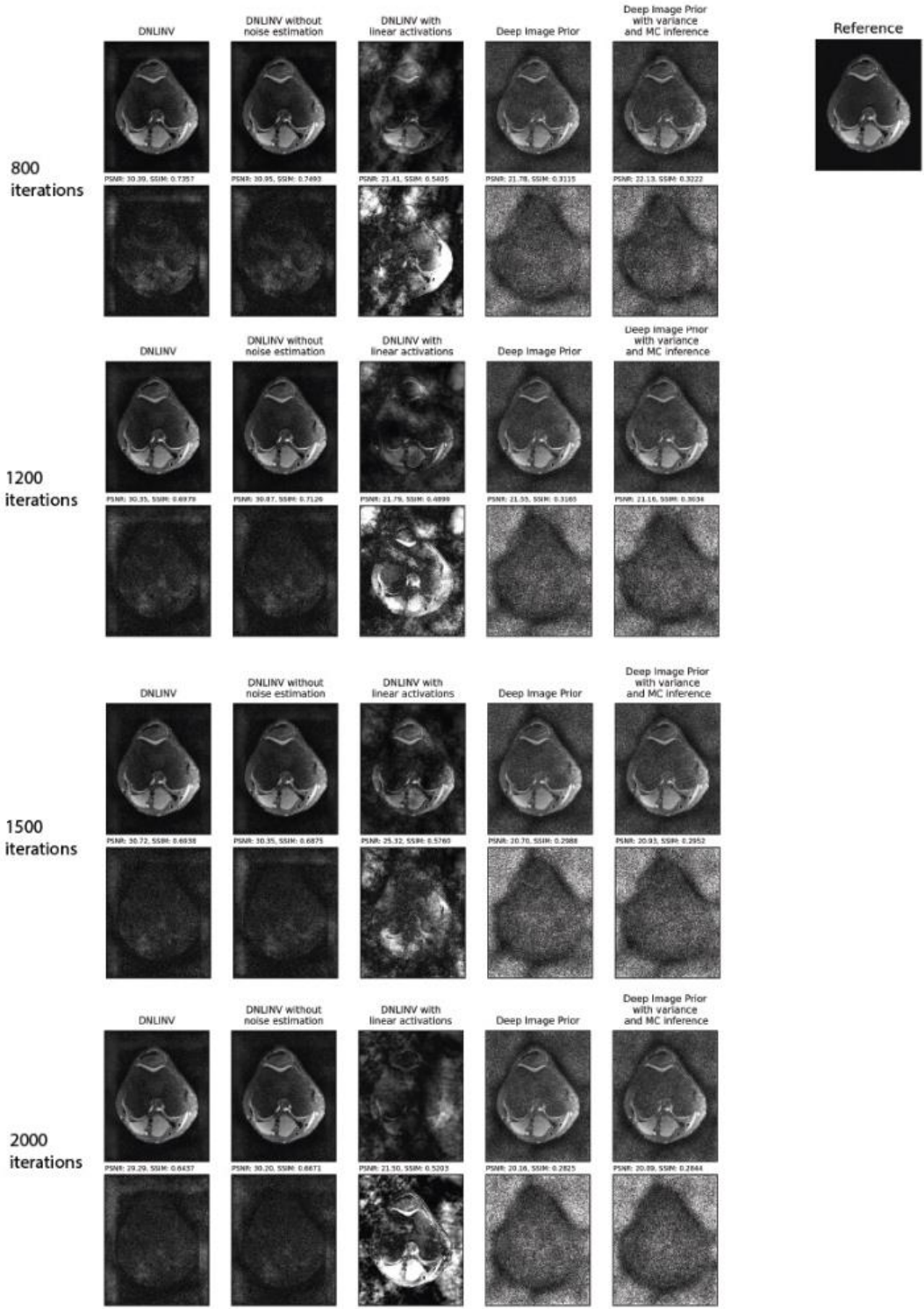

Supp. Fig. 10: Ablation experiment results across iterations. 1500 iterations was chosen as the approximate optimal number of iterations for DNLINV. Within the first few hundred iterations, the DIP methods overfit to noise, whereas the DNLINV methods only show overfitting to noise at much later iterations. DNLINV with early-stopping allows for a high quality result without overfitting to noise.

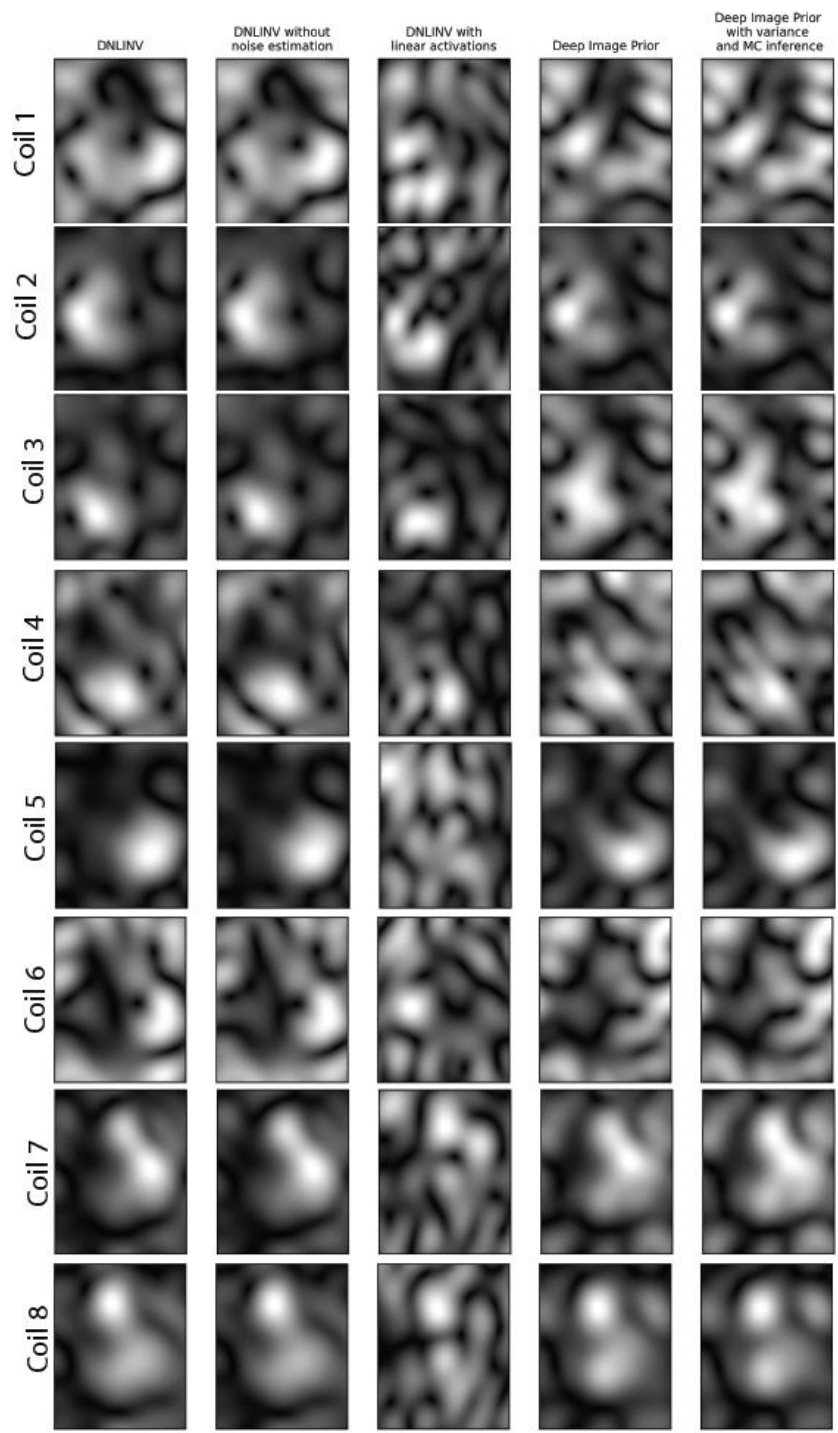

Supp Fig 11: Coil sensitivity maps estimated in the knee ablation experiment. DNLINV and DNLINV without noise estimation provide the most reasonable coil sensitivity maps, whereas the other methods show both higher noise levels as well as less realistic coil sensitivity profiles.